\pdfoutput=1
\documentclass[acmsmall,screen,nonacm]{acmart}

\usepackage[ruled,lined,linesnumbered,vlined]{algorithm2e}

\SetCommentSty{mycommfont}
\DontPrintSemicolon % Do not show the semicolons.

\SetKwProg{Fn}{Function}{:}{}

\usepackage{amsmath}
\usepackage{amsfonts}
\usepackage{amsthm}
\usepackage{balance}
\usepackage{enumitem}
\usepackage{graphicx}
\usepackage{listings}
\usepackage{multicol}
\usepackage{multirow}
\usepackage{subcaption}
\usepackage{url}
\usepackage{xspace}
\usepackage{xcolor}
\usepackage{tcolorbox}
\usepackage{colortbl}
\usepackage{siunitx}
\newcommand{\mypercent}[1]{\SI{#1}{\percent}}

\usepackage{tikz}
\usepackage{pgfplots}
\newcommand*\circled[1]{\tikz[baseline=(char.base)]{%
    \node[shape=circle,draw,inner sep=0.5pt,
          minimum size=0pt,minimum height=0pt,minimum width=0pt]
      (char) {\upshape\fontsize{0.85em}{0.85em}\selectfont#1};}}
\usetikzlibrary{shapes}
\usetikzlibrary{shapes.geometric}
\usetikzlibrary{arrows.meta, positioning}

\newcommand{\eg}{\mbox{\textit{e.g.}}\xspace}

\newcommand{\ie}{\mbox{\textit{i.e.}}\xspace}
\definecolor{BlindColorTolOne}{HTML}{332288}
\definecolor{BlindColorTolTwo}{HTML}{117733} % green
\definecolor{BlindColorTolThree}{HTML}{44AA99}
\definecolor{BlindColorTolFour}{HTML}{88CCEE}
\definecolor{BlindColorTolFive}{HTML}{DDCC77}
\definecolor{BlindColorTolSix}{HTML}{CC6677} % light red
\definecolor{BlindColorTolSeven}{HTML}{AA4499}
\definecolor{BlindColorTolEight}{HTML}{882255}

\definecolor{BlindColorWongOne}{HTML}{000000} % black
\definecolor{BlindColorWongTwo}{HTML}{E69F00}
\definecolor{BlindColorWongThree}{HTML}{56B4E9}
\definecolor{BlindColorWongFour}{HTML}{009E73}
\definecolor{BlindColorWongFive}{HTML}{F0E442}
\definecolor{BlindColorWongSix}{HTML}{0072B2} % blue
\definecolor{BlindColorWongSeven}{HTML}{D55E00}
\definecolor{BlindColorWongEight}{HTML}{CC79A7}
\newif\ifshowcomments
\showcommentstrue

\definecolor{mygreen}{HTML}{02818a}

\mathchardef\mhyphen="2D

\newcounter{FindingCounter}
\newcommand{\finding}[1]{
	\begin{tcolorbox}
		\textbf{RQ\refstepcounter{FindingCounter}\theFindingCounter}: #1
	\end{tcolorbox}
}

\newcommand{\myparagraph}[1]{
  \vspace*{0.04cm}
  \noindent \textit{\textbf{#1.}}\quad
}

\newcommand{\itemParagraph}[1]{%
  \par\addvspace{2pt plus 1pt}%
  \noindent\textit{#1.}\quad\ignorespaces}

\newcommand{\projectname}[1]{\mbox{\textsf{#1}}\xspace} % \xspace is not needed here.
\SetKwData{AlgTrue}{true}
\SetKwData{AlgFalse}{false}
\SetKw{AlgBreak}{break}
\SetKw{AlgContinue}{continue}

\newcounter{myUniqueIdCounter}

\makeatletter
\newcommand{\myGetOrAssignID}[1]{%
  \ifcsname myMap@#1\endcsname%
    \csname myMap@#1\endcsname%
  \else%
    \stepcounter{myUniqueIdCounter}%
    \expandafter\xdef\csname myMap@#1\endcsname{\themyUniqueIdCounter}%
    \themyUniqueIdCounter%
  \fi%
}
\makeatother

\newcommand{\anonymizedId}[1]{\ifx\useAnonymizedId\undefined%
  #1%
\else%
  \myGetOrAssignID{#1}%
\fi}

\usepackage{mathpartir}
\usepackage{wrapfig}
\usepackage{stmaryrd}
\definecolor{mpirKw}{HTML}{0072B2}
\definecolor{mpirCons}{HTML}{D55E00}
\definecolor{mpirMeta}{HTML}{009E73}
\definecolor{mpirComment}{HTML}{6E6E6E}
\definecolor{mpirFlag}{HTML}{CC79A7}

\lstdefinelanguage{mpir}{
  morekeywords={add,sub,mul,udiv,sdiv,urem,srem,shl,lshr,ashr,and,or,xor,
    icmp,fcmp,select,zext,sext,trunc,call,phi,freeze,br,ret,store,load,
    getelementptr,extractvalue,insertvalue,fneg,fadd,fsub,fmul,fdiv,
    match,metaflag,constraint,to,label,ptr,
    neg,addLike,shift,shr,zextOrSext,selectEqZero,signum,logicalAnd,
    logicalOr,logicalOp,logicalShift,bitwiseLogic,idiv,irem,maxOrMin,
    selectMinMax,signBitCheck,sextLike,itofp,fptoi,ptrToIntOrAddr,fnegLike,
    selectEq},
  morekeywords=[2]{nsw,nuw,exact,disjoint,nneg,samesign,ne,eq,ult,ugt,slt,sgt},
  morekeywords=[3]{Unary,Binary,Nary,Cast},
  sensitive=true,
  morecomment=[l][\color{mpirCons}\bfseries]{;;},
  morecomment=[l][\color{mpirComment}\itshape]{;\ },
  morestring=[b]",
  alsoletter={?\%.\$_},
  alsodigit={-},
}

\lstdefinestyle{mpirstyle}{
  language=mpir,
  basicstyle=\footnotesize\ttfamily,
  keywordstyle=\color{mpirKw}\bfseries,
  keywordstyle=[2]\color{mpirFlag},
  keywordstyle=[3]\color{mpirKw}\itshape,
  stringstyle=\color{mpirCons},
  commentstyle=\color{mpirComment}\itshape,
  columns=fullflexible,
  keepspaces=true,
  showstringspaces=false,
  escapechar=@,
  xleftmargin=1em,
  numbers=none,
  aboveskip=2pt,
  belowskip=2pt,
}

\definecolor{cppMatch}{HTML}{009E73}

\lstdefinestyle{cppstyle}{
  language=C++,
  emph={match,m_OneUse,m_UDiv,m_Value,m_c_Add,m_ZExt,m_SpecificICmp,m_And,
    m_LowBitMask,m_ZeroInt,m_ZExtOrSelf,m_LShr,m_Deferred,m_APInt,
    willNotOverflowUnsignedMul,willNotOverflowUnsignedAdd,popcount,
    m_Specific,m_Add,m_NSWAdd,m_Power2,m_AddLike,m_Shift,m_Signum,
    isKnownNonZero,hasNoSignedWrap},
  emphstyle=\color{cppMatch}\bfseries,
  basicstyle=\footnotesize\ttfamily,
  keywordstyle=\color{mpirKw}\bfseries,
  commentstyle=\color{mpirComment}\itshape,
  stringstyle=\color{mpirCons},
  columns=fullflexible,
  keepspaces=true,
  showstringspaces=false,
  escapechar=@,
  xleftmargin=1em,
  numbers=none,
  aboveskip=2pt,
  belowskip=2pt,
}

\lstdefinestyle{llvmstyle}{
  language=mpir,
  basicstyle=\footnotesize\ttfamily,
  keywordstyle=\color{mpirKw}\bfseries,
  keywordstyle=[2]\color{mpirFlag},
  commentstyle=\color{mpirComment}\itshape,
  morekeywords={define},
  columns=fullflexible,
  keepspaces=true,
  escapechar=@,
  xleftmargin=1em,
  aboveskip=2pt,
  belowskip=2pt,
}

\definecolor{hlChange}{HTML}{FFE08A}
\newcommand{\hlc}[1]{{\setlength{\fboxsep}{0.6pt}\colorbox{hlChange}{#1}}}
\newcommand{\hmv}[1]{\hlc{\hcode{#1}}}
\newcommand{\hsp}{\texttt{\ }}
\definecolor{hlPrecond}{HTML}{CDE7F7}
\newcommand{\hlp}[1]{{\setlength{\fboxsep}{0.6pt}\colorbox{hlPrecond}{#1}}}

\newcommand{\fc}[1]{\lstinline[style=mpirstyle,basicstyle=\scriptsize\ttfamily]{#1}}

\newcommand{\hcode}[1]{\mbox{\lstinline[style=mpirstyle,basicstyle=\ttfamily]{#1}}}
\newcommand{\hcpp}[1]{\mbox{\lstinline[style=cppstyle,basicstyle=\ttfamily]{#1}}}

\newcommand{\code}[1]{\mbox{\lstinline[style=mpirstyle,basicstyle=\small\ttfamily,keywordstyle=\relax,keywordstyle={[2]\relax},keywordstyle={[3]\relax},commentstyle=\relax,morecomment={[l][\relax]{;;}},stringstyle=\relax]{#1}}}
\newcommand{\cpp}[1]{\mbox{\lstinline[style=cppstyle,basicstyle=\small\ttfamily,keywordstyle=\relax,stringstyle=\relax,emphstyle=\relax]{#1}}}

\newcommand{\pre}{\ensuremath{\mathit{pre}}}
\newcommand{\src}{\ensuremath{\mathit{src}}}
\newcommand{\tgt}{\ensuremath{\mathit{tgt}}}
\newcommand{\MVars}{\mathcal{X}}
\newcommand{\LNames}{\mathcal{L}}
\newcommand{\Consts}{\mathcal{K}}
\newcommand{\Ops}{\mathcal{F}}
\newcommand{\FlagSet}{\mathit{Flag}}
\newcommand{\fl}{\mathit{fl}}
\newcommand{\KeySet}{\mathit{Key}}
\newcommand{\mkc}{\kw{c}}
\newcommand{\mko}[1]{\kw{o}(#1)}
\newcommand{\kwas}{\kw{as}}
\newcommand{\kwin}{\kw{in}}
\newcommand{\res}{\kw{res}}
\newcommand{\kpred}{\kw{pred}}
\newcommand{\dom}{\mathrm{dom}}
\newcommand{\srt}{\varsigma}
\newcommand{\subsort}{\sqsubseteq}
\newcommand{\coerce}{\preceq}
\newcommand{\cto}[2]{#1{\downarrow}_{#2}}
\newcommand{\sOpd}{\kw{Opd}}
\newcommand{\sVal}{\kw{Val}}
\newcommand{\sInst}{\kw{Inst}}
\newcommand{\sFn}{\kw{Fn}}
\newcommand{\sBB}{\kw{BB}}
\newcommand{\sNat}{\kw{Nat}}
\newcommand{\sBool}{\kw{Bool}}
\newcommand{\sInt}{\kw{Int}}
\newcommand{\sStr}{\kw{Str}}
\newcommand{\sPred}{\kw{Pred}}
\newcommand{\sIntr}{\kw{Intr}}
\newcommand{\sType}{\kw{Type}}
\newcommand{\sBV}{\kw{BV}}
\newcommand{\mv}{x}
\newcommand{\mvi}{x_{I}}% bound by \kwas to an instruction
\newcommand{\mvbb}{x_{\mathit{bb}}}% bound by \kwin to a basic block
\newcommand{\lv}{\ell}
\newcommand{\op}{\mathit{op}}

\newcommand{\kw}[1]{\ensuremath{\mathsf{#1}}}
\newcommand{\fn}[1]{\mathsf{#1}}
\newcommand{\flagspec}{\varphi}
\newcommand{\mks}{\mu}
\newcommand{\pat}{\bar{P}}
\newcommand{\varnt}{P}

\newcommand{\Lower}{\fn{Lower}}
\newcommand{\Match}{\fn{Match}}
\newcommand{\Struct}{\Match_{\fn{struct}}}

\newcommand{\opcode}{\fn{opc}}
\newcommand{\opnds}{\fn{ops}}
\newcommand{\flagsof}{\fn{flags}}
\newcommand{\typeof}{\fn{ty}}
\newcommand{\tyarg}{\fn{tyarg}}
\newcommand{\keysof}{\fn{keys}}
\newcommand{\keyspec}{\psi}
\newcommand{\blockof}{\fn{blk}}
\newcommand{\ValsOf}{\fn{Vals}}
\newcommand{\FactsOf}{\fn{Facts}}

\newcommand{\Req}{\fn{Req}}
\newcommand{\JoinSig}{\fn{Typings}}

\newcommand{\SitesOf}{\fn{sites}}

\newcommand{\llvmPRURL}[1]{https://github.com/llvm/llvm-project/pull/#1}
\newcommand{\llvmPR}[1]{\href{\llvmPRURL{#1}}{PR~#1}}
\newcommand{\llvmPRNum}[1]{\href{\llvmPRURL{#1}}{#1}}
\newcommand{\llvmIssueURL}[1]{https://github.com/llvm/llvm-project/issues/#1}
\newcommand{\llvmIssue}[1]{\href{\llvmIssueURL{#1}}{Issue~#1}}

\newcommand{\mulop}{\mathbin{\cdot}}
\newcommand{\udivop}{\mathbin{/_u}}

\newcommand{\shrop}{\mathbin{\gg}}

\newcommand{\sem}[1]{\llbracket #1 \rrbracket}
\newcommand{\asem}[1]{\widehat{\llbracket #1 \rrbracket}}
\newcommand{\err}{\kw{err}}
\newcommand{\conc}[1]{\kw{const}(#1)}
\newcommand{\sitesv}[1]{\kw{at}(#1)}
\newcommand{\unk}{\top}
\newcommand{\AVal}{\hat{\mathcal{D}}}
\newcommand{\CDom}{\mathcal{D}}
\newcommand{\CVal}{\mathbb{C}}
\newcommand{\gam}{\gamma}
\newcommand{\feat}{\alpha(F)}
\newcommand{\true}{\kw{true}}
\newcommand{\false}{\kw{false}}

\newcommand{\matches}{\rightsquigarrow}
\newcommand{\fsat}{\models}
\newcommand{\pmatch}[3]{#1 \vdash #2 \matches #3}

\AtBeginEnvironment{lemma}{\crefalias{theorem}{lemma}}
\AtBeginEnvironment{corollary}{\crefalias{theorem}{corollary}}
\AtBeginEnvironment{definition}{\crefalias{theorem}{definition}}

\theoremstyle{acmplain}
\newcommand{\restatename}{}
\newtheorem*{restate*}{\restatename}
\newenvironment{restate}[2]
  {\renewcommand{\restatename}{\Cref{#1}}\begin{restate*}[#2]}
  {\end{restate*}}

\showcommentsfalse

\newcommand{\sys}{\projectname{MetaphIR}}
\newcommand{\llvm}{LLVM}
\newcommand{\llvmir}{LLVM~IR}
\newcommand{\instcombine}{InstCombine}
\newcommand{\instsimplify}{InstSimplify}
\newcommand{\patternmatch}{\texttt{PatternMatch.h}}
\newcommand{\valuetracking}{\texttt{ValueTracking}}
\newcommand{\daggrep}{\textsf{daggrep}}
\newcommand{\Daggrep}{\textsf{Daggrep}}
\newcommand{\corpusName}{\textsf{llvm-opt-benchmark}}
\newcommand{\alive}{Alive}
\newcommand{\aliveTwo}{Alive2}

\newcommand{\stdlib}{\texttt{stdlib.mpir}}

\newcommand{\metavar}{meta variable}
\newcommand{\metavars}{meta variables}

\newcommand{\metaflag}{metaflag}
\newcommand{\metaflags}{metaflags}
\newcommand{\composite}{composite matcher}
\newcommand{\composites}{composite matchers}
\newcommand{\variant}{core pattern}
\newcommand{\variants}{core patterns}
\newcommand{\NumStagesWord}{five}
\newcommand{\StageLower}{Stage~\circled{1}}
\newcommand{\StageLookup}{Stage~\circled{2}}
\newcommand{\StageFilter}{Stage~\circled{3}}
\newcommand{\StageMatch}{Stage~\circled{4}}
\newcommand{\StageEval}{Stage~\circled{5}}
\newcommand{\instantiation}{instantiation}
\newcommand{\instantiations}{instantiations}
\newcommand{\site}{site}
\newcommand{\sites}{sites}
\newcommand{\fact}{fact}
\newcommand{\facts}{facts}
\newcommand{\measure}{measure}

\newcommand{\features}{features}
\newcommand{\abstractFace}{abstract transformer}
\newcommand{\Metaflags}{Metaflags}

\newcommand{\Variant}{Core Pattern}
\newcommand{\Composite}{Composite matcher}

\newcommand{\Instantiations}{Instantiations}
\newcommand{\Site}{Site}
\newcommand{\Sites}{Sites}
\newcommand{\Facts}{Facts}

\newcommand{\Features}{Features}
\newcommand{\feature}{feature}

\newcommand{\abstractFaces}{abstract transformers}
\newcommand{\TMetavar}{Meta Variable}
\newcommand{\TMetavars}{Meta Variables}
\newcommand{\TComposites}{Composite Matchers}
\newcommand{\TAbstractFaces}{Abstract Transformers}

\newcommand{\gt}{ground truth}
\newcommand{\dsl}{DSL}

\newcommand{\InstCombineLocOld}{36,000}
\newcommand{\InstCombineLocNew}{50,000}
\newcommand{\LlvmVerOld}{12}
\newcommand{\LlvmVerNew}{22}

\newcommand{\NumInstMatchers}{49}
\newcommand{\NumOperators}{105}
\newcommand{\NumValueTrackOps}{21}
\newcommand{\MaxTypings}{256}
\newcommand{\NumConstClasses}{23}
\newcommand{\NumStdComposites}{23}

\newcommand{\WasmSizeMB}{4.2}

\newcommand{\NumPRsWord}{twenty}
\newcommand{\NumPatterns}{23}
\newcommand{\NumPatternInsts}{75}
\newcommand{\NumPatternConstraints}{20}
\newcommand{\MaxPatternInsts}{six}

\newcommand{\NumBenchModules}{68,843}
\newcommand{\NumTestFiles}{1,912}
\newcommand{\NumProjects}{216}
\newcommand{\NumFunctions}{4,771,730}
\newcommand{\NumFunctionsShort}{4.8M}
\newcommand{\NumFunctionsApprox}{4.8~million}% corpus alone; the test suites add a negligible share
\newcommand{\CorpusSizeGB}{11.9}
\newcommand{\IndexSizeGB}{2.70}
\newcommand{\WebIndexSizeMB}{1,238}
\newcommand{\PacksSizeGB}{5.56}

\newcommand{\NumExactPRsWord}{twenty}
\newcommand{\NumZeroGTPRsWord}{six}% folds with |GT| = |P| = 0, counted as exact
\newcommand{\DaggrepZeroGTP}{68,175}% daggrep |P| for pr209322, whose |GT| is 0
\newcommand{\RunPatternBenchMatches}{18}

\newcommand{\PRMergedFrom}{June 20}
\newcommand{\PRMergedTo}{August 18, 2026}

\newcommand{\MainPrecision}{\mypercent{100}}

\newcommand{\MainFP}{0}

\newcommand{\TestPrecision}{\mypercent{100.00}}
\newcommand{\TestRecall}{\mypercent{99.29}}
\newcommand{\TestG}{141}% |GT|

\newcommand{\TestFN}{one}% pr214376, a non-splat vector constant
\newcommand{\PipeMacroPrecision}{\mypercent{70.54}}
\newcommand{\PipeMacroRecall}{\mypercent{79.56}}
\newcommand{\MaxFuncSurvive}{\mypercent{0.38}}
\newcommand{\RunPatternFuncs}{396}
\newcommand{\RunPatternMods}{223}

\newcommand{\IndexedSec}{0.63}
\newcommand{\IndexBuildSec}{142}% wall clock, 80 threads (REPORT.md)
\newcommand{\IndexBuildNoVFSec}{113}
\newcommand{\NumThreads}{80}
\newcommand{\MedFuncSurvive}{893}
\newcommand{\MedFuncSurvivePct}{\mypercent{0.019}}
\newcommand{\MaxFuncSurviveAbs}{17,996}

\newcommand{\MedModSurvive}{361}
\newcommand{\MedModSurvivePct}{\mypercent{0.51}}

\newcommand{\IndexedSecRangeLo}{0.63}
\newcommand{\IndexedSecRangeHi}{3.26}
\newcommand{\IndexedSecMedian}{1.27}
\newcommand{\FullScanSecMedian}{26.2}
\newcommand{\NightlyMinMedian}{47}
\newcommand{\MedIndexSpeedup}{22$\times$}% median of the per-PR ratios
\newcommand{\MinIndexSpeedup}{8.0$\times$}
\newcommand{\MaxIndexSpeedup}{57$\times$}
\newcommand{\MedNoConstrFuncs}{2,167}
\newcommand{\MedNoConstrFactor}{2.4$\times$}% MedNoConstrFuncs / MedFuncSurvive
\newcommand{\WebFullSecMedian}{42.3}% per PR: per-pattern medians, summed
\newcommand{\WebNoConstrSecMedian}{54.7}

\newcommand{\WebNoConstrFactorMax}{4.3$\times$}% \llvmPR{209322}
\newcommand{\WebNoConstrSecMax}{383}% \llvmPR{209322}
\newcommand{\WebFullSecMax}{90}% \llvmPR{209322}
\newcommand{\WebFullCandsMax}{19}% \llvmPR{209322}
\newcommand{\WebNoConstrCandsMax}{68,782}% \llvmPR{209322}
\newcommand{\WebFullMBMax}{852}% \llvmPR{209322}
\newcommand{\WebNoConstrMBMax}{3,564}% \llvmPR{209322}

\newcommand{\MedSecNoConstr}{1.65}
\newcommand{\NumNoConstrWorse}{20}% patterns that keep more functions without constraint prefiltering

\newcommand{\LlvmCommit}{79b321b}
\newcommand{\NumPMUsed}{206}
\newcommand{\NumPMCovered}{178}
\newcommand{\PMCoveredPct}{\mypercent{86.4}}
\newcommand{\NumPMCalls}{7,022}
\newcommand{\PMCallsCoveredPct}{\mypercent{96.4}}
\newcommand{\NumVTUsed}{60}
\newcommand{\NumVTCovered}{24}
\newcommand{\VTCoveredPct}{\mypercent{40.0}}
\newcommand{\NumVTCalls}{437}
\newcommand{\VTCallsCoveredPct}{\mypercent{67.0}}
\newcommand{\NumMatchExprs}{2,359}% match(V, ...) calls in InstCombine
\newcommand{\NumMatchExprsCovered}{2,129}% ... that call only covered matchers
\newcommand{\MatchExprsCoveredPct}{\mypercent{90.3}}
\newcommand{\DaggrepExpressible}{3}
\newcommand{\DaggrepInsts}{118}% instruction lines of the daggrep IR patterns
\newcommand{\DaggrepPrecision}{\mypercent{4.29}}
\newcommand{\DaggrepRecall}{\mypercent{100.00}}
\newcommand{\DaggrepFP}{86,123}
\newcommand{\DaggrepTestPrecision}{\mypercent{17.10}}
\newcommand{\DaggrepSecMedian}{32.5}
\newcommand{\DaggrepSecLo}{20.7}
\newcommand{\DaggrepSecHi}{86.7}

\usepackage{cleveref} % This package must be loaded in the end.

\Crefname{algocf}{Algorithm}{Algorithms}
\crefname{algocf}{Algorithm}{Algorithms}

\Crefname{algorithm}{Algorithm}{Algorithms}
\crefname{algorithm}{Algorithm}{Algorithms}

\crefname{appendix}{Appendix}{Appendices}
\Crefname{appendix}{Appendix}{Appendices}

\Crefname{figure}{Figure}{Figures}
\crefname{figure}{Figure}{Figures}

\crefname{listing}{Listing}{Listings}
\Crefname{listing}{Listing}{Listings}

\Crefname{table}{Table}{Tables}
\crefname{table}{Table}{Tables}

\crefname{thm}{Theorem}{Theorems}
\Crefname{thm}{Theorem}{Theorems}

\crefname{definition}{Definition}{Definitions}
\Crefname{definition}{Definition}{Definitions}

\crefname{lemma}{Lemma}{Lemmas}
\Crefname{lemma}{Lemma}{Lemmas}

\crefname{corollary}{Corollary}{Corollaries}
\Crefname{corollary}{Corollary}{Corollaries}

\crefname{proof}{Proof}{Proofs}
\Crefname{proof}{Proof}{Proofs}

\crefname{equation}{Equation}{Equations}
\Crefname{equation}{Equation}{Equations}

\crefformat{chapter}{\S~#2#1#3}
\crefmultiformat{chapter}{\S\S~#2#1#3}{ and~#2#1#3}{, #2#1#3}{, and~#2#1#3}

\crefformat{section}{\S~#2#1#3}
\crefmultiformat{section}{\S\S~#2#1#3}{ and~#2#1#3}{, #2#1#3}{, and~#2#1#3}

\renewcommand{\myparagraph}[1]{%
  \par\addvspace{2pt plus 1pt}%
  \noindent\textit{\textbf{#1.}}\quad\ignorespaces
}

\renewcommand{\finding}[1]{%
  \begin{tcolorbox}[boxrule=0.5pt,boxsep=1.5pt,left=3pt,right=3pt,top=3pt,bottom=3pt,arc=2pt]
    \textbf{RQ\refstepcounter{FindingCounter}\theFindingCounter}: #1
  \end{tcolorbox}%
}

\makeatletter
\if@ACM@anonymous
  \def\useAnonymizedId{}
\fi
\makeatother

\pgfplotsset{compat=1.18}

\begin{document}

\title{\sys{}: Corpus-Scale Applicability Analysis for LLVM Peephole Optimizations}

% Authors shared by main.tex and supplement.tex.

\author{Hongxu Xu}
\orcid{0009-0003-0926-220X}
\affiliation{%
  \institution{University of Waterloo}
  \city{Waterloo}
  \country{Canada}
}
\email{hongxu.xu@uwaterloo.ca}

\author{Xintong Zhou}
\orcid{0009-0002-6444-5431}
\affiliation{%
  \institution{University of Waterloo}
  \city{Waterloo}
  \country{Canada}
}
\email{x27zhou@uwaterloo.ca}

\author{Chunhao Liao}
\orcid{0009-0004-4379-9454}
\affiliation{%
  \institution{University of Waterloo}
  \city{Waterloo}
  \country{Canada}
}
\email{chunhao.liao@uwaterloo.ca}

\author{Chengnian Sun}
\orcid{0000-0002-0862-2491}
\affiliation{%
  \institution{University of Waterloo}
  \city{Waterloo}
  \country{Canada}
}
\email{cnsun@uwaterloo.ca}

\begin{abstract}
Peephole optimizations replace short \emph{source} instruction sequences
with more efficient equivalents, typically guarded by \emph{preconditions}
such as requiring a constant operand to be a power of two.
Since every new rewrite adds maintenance burden and compile-time cost,
compiler developers need to know how often real-world programs
both match the source pattern and satisfy its preconditions
before committing to an implementation.
Existing workflows fall short:
structural search over an intermediate representation (IR) corpus
of \NumFunctionsApprox{} functions
ignores preconditions and therefore overcounts,
whereas implementing the rewrite and recompiling the corpus
yields exact counts only after paying the engineering cost,
with each run of the corpus's on-demand recompilation pipeline taking a median of
\NightlyMinMedian{} minutes.
This delay is especially limiting when generalizing a rewrite,
as developers often iterate over multiple variants to trade off
coverage against complexity.

We introduce \sys{}, a system pairing a domain-specific language
for expressing \llvm{} peephole optimizations
with an indexed search engine over large IR corpora.
Using a syntax modeled on \llvm{}'s C++ matchers,
\sys{} patterns augment \llvm{} IR with holes
and Boolean preconditions over matched values.
To answer queries at interactive speed,
\sys{} builds compact per-function summaries ahead of time
and uses abstract interpretation over those summaries
to soundly filter out functions that cannot satisfy a query's constraints.
Across \NumPRsWord{} recent peephole optimizations
from \llvm{}'s \instcombine{} pass,
\sys{} recovers the exact set of functions where each rewrite applies in isolation,
achieving \MainPrecision{} precision and recall.
Queries complete in a median of \IndexedSecMedian{} seconds---a
\MedIndexSpeedup{} speedup over a full corpus scan---enabling
developers to quickly gauge an optimization's reach and generality
prior to implementation.
\end{abstract}




\maketitle

\section{Introduction}
\label{sec:intro}

Peephole optimizations replace short instruction sequences with
more efficient ones~\cite{mckeeman_peephole_1965}.
In \llvm{}~\cite{lattner_llvm_2004},
most of these optimizations are implemented
in the \instcombine{} pass~\cite{llvm_instcombine},
which encodes thousands of them in C++ as \emph{rewrites}.
A rewrite $\pre{} \vdash \src{} \rightarrow \tgt{}$
replaces instructions that match a \emph{source} (\src{})
with a \emph{target} (\tgt{}),
provided that a \emph{precondition} (\pre{}) holds.
For example, one rewrite is
\(
  \mathit{isPowerOf2}(C) \vdash x / C \rightarrow x \gg \log_2 C,
\)
which transforms an unsigned division by a power of two into a right shift.
In C++, a rewrite matches \src{} and checks \pre{}
with the \patternmatch{} matcher library~\cite{llvm_patternmatch},
such as \cpp{m_Power2} for $C$,
and then builds \tgt{}.
\instcombine{} keeps growing:
it has expanded from about \InstCombineLocOld{} lines of code in
\llvm{}~\LlvmVerOld{} to over \InstCombineLocNew{} in \llvm{}~\LlvmVerNew{}.
Developers add new rewrites or generalize existing ones
to capture missed optimizations,
and automated techniques now discover such optimizations
at scale~\cite{sasnauskas_souper_2018,mukherjee_hydra_2024,xu_lpo_2026}.

Each new or generalized rewrite raises the same question: \emph{does it matter?}
A rewrite is worth its maintenance and compile-time overhead only
if real-world code frequently matches its \src{}
and satisfies its \pre{}.
To answer this, developers search a corpus of
intermediate representation (IR) compiled from real-world programs,
such as \corpusName{}~\cite{llvm_opt_benchmark},
which contains \NumFunctionsApprox{} functions.
In practice, developers use \daggrep{}~\cite{daggrep,daggrep_issue},
which takes a concrete IR instance of \src{} as a query
and finds instructions with the same structure across the corpus.
However, \daggrep{} relies purely on structural matching
and cannot express preconditions (\pre{}).
For example, to search for the division rewrite above,
a structural query can only match division by any constant divisor;
ignoring the precondition $\mathit{isPowerOf2}(C)$,
\daggrep{} overmatches on instructions such as $x / 3$
where the rewrite cannot apply.
Its applicability counts are therefore inaccurate.
Alternatively, developers can implement the rewrite in C++
and recompile the corpus.
While exact, this measures applicability only after paying
the implementation cost,
and recompiling for each candidate rewrite is slow.
\corpusName{} provides an on-demand recompilation pipeline,
and each run takes a median of \NightlyMinMedian{}
minutes~\cite{llvm_opt_benchmark_nightly_actions}.
This latency makes iteratively exploring variations of a rewrite
impractical.

\myparagraph{Challenges}
Accurately and quickly assessing where a rewrite applies poses two challenges.
First, queries must express both \src{} and \pre{} in a concise, declarative
notation that developers can easily write.
This is difficult because rewrites in \instcombine{} are not simple syntactic
templates:
\src{} abstracts over types, bit widths, instruction flags (\eg{}, \code{nsw}),
operand commutativity, and interchangeable opcodes (\eg{}, accepting either
\code{add} or \code{or disjoint}),
while \pre{} involves semantic checks normally implemented in C++,
such as single-use restrictions, constant predicates like powers of two,
and dataflow facts like absence of overflow.
Second,
query execution must be fast enough for developers to assess
and refine candidate rewrites interactively across millions of functions.
Scanning every function in the corpus is impractical because loading
and parsing IR bitcode dominates query time.
Achieving interactive speed therefore requires an index that prunes irrelevant
functions without reading their IR.
However, designing such an index is nontrivial:
to prune candidates effectively,
it must exploit both structural patterns and semantic preconditions,
yet never prune a function that contains a match.

\begin{wrapfigure}[4]{r}{0.32\linewidth}
  \vspace{-1.2em}
  \centering
  \begin{tikzpicture}[
    hl/.style={rounded corners=1.2pt, inner xsep=1.2pt, inner ysep=1.2pt,
    anchor=base},
    lbl/.style={font=\scriptsize, inner sep=0pt, align=center},
    arr/.style={-{Stealth[length=1.2mm, width=1.2mm]}, thin, black!60},
  ]
    \node[anchor=base west, inner sep=0pt, font=\footnotesize\ttfamily] (op) at (0,0) {\textcolor{mpirKw}{\textbf{udiv}}\ i};
    \node[hl, fill=hlChange, xshift=0.5pt] (bits) at (op.base east) [anchor=base west] {\footnotesize\ttfamily ?bits};
    \node[anchor=base west, inner sep=0pt, font=\footnotesize\ttfamily] (sp1) at (bits.base east) {\ };
    \node[hl, fill=hlChange] (x) at (sp1.base east) [anchor=base west] {\footnotesize\ttfamily ?x};
    \node[anchor=base west, inner sep=0pt, font=\footnotesize\ttfamily] (comma) at (x.base east) {,\ };
    \node[hl, fill=hlChange] (c) at (comma.base east) [anchor=base west] {\footnotesize\ttfamily ?C};
    \node[hl, fill=hlPrecond, xshift=1pt] (flag) at (c.base east) [anchor=base west] {\footnotesize\ttfamily :power2};

    % Annotations
    \node[lbl, below=3.2mm of x.south, xshift=-2.5mm] (holes_lbl) {\metavars{} (holes)};
    \draw[arr] ([xshift=-3mm]holes_lbl.north) -- ([yshift=-0.5pt]bits.south);
    \draw[arr] ([xshift=2.5mm]holes_lbl.north) -- ([yshift=-0.5pt]x.south);
    \draw[arr] ([xshift=5mm]holes_lbl.north) -- ([yshift=-0.5pt]c.south);

    \node[lbl, anchor=base west] (flag_lbl) at (flag.base |- holes_lbl.base) {\metaflag{}};
    \draw[arr] (flag_lbl.north) -- ([yshift=-0.5pt]flag.south);
  \end{tikzpicture}
  \vspace{-0.8em}
  \caption{A \sys{} Pattern.}
  \Description{A MetaphIR pattern udiv i?bits ?x, ?C:power2 with ?bits, ?x,
  and ?C highlighted as meta variables (holes)
  and :power2 highlighted as a metaflag.}
  \label{fig:intro-pattern}
\end{wrapfigure}
\myparagraph{\sys{}}
To address these challenges, we present \sys{}, a system that pairs a
domain-specific language (\dsl{}) for specifying the source pattern and
precondition of \llvm{} peephole optimizations with an indexed search engine
that accurately finds all matches across a large corpus in seconds.
First, to achieve expressiveness (Challenge~1),
a \sys{} \emph{pattern} (\eg{}, \cref{fig:intro-pattern}) describes both \src{}
and \pre{} of a rewrite:
\src{} is \llvmir{} with holes (\metavars{}, highlighted in \hlc{yellow}),
which \sys{} matches structurally,
and \pre{} is a set of \emph{constraints} on the values that the holes bind.
To make patterns intuitive to write, \sys{} directly mirrors \llvm{}'s C++
matchers
and analysis APIs:
common checks can be attached directly to holes as \emph{\metaflags{}}
(\eg{}, \hlp{\code{:power2}} on \code{?C} in \cref{fig:intro-pattern}
corresponds to \llvm{}'s \cpp{m_Power2} matcher).
Because \sys{} reuses the checks developers already know,
they can easily turn a concrete IR example into a pattern
by replacing specific values with holes and attaching constraints.
Second, to achieve interactive speed at scale (Challenge~2),
\sys{} prunes irrelevant functions without loading their IR:
it precomputes and indexes a compact summary of each function's
\emph{\features{}},
which instructions occur and which \facts{} hold in each operand slot.
A single constraint evaluator operates over two domains:
it checks concrete matches on the IR,
and,
when evaluated over the indexed \features{},
it acts as an abstract interpretation~\cite{cousot_abstract_1977} of the
constraint language
that decides whether any match in the function is possible.
We prove that this abstract evaluation is sound, \ie{},
it never skips a function
containing a valid match.

We evaluate \sys{} on \NumPRsWord{} recent \llvm{} rewrites
against \corpusName{}~\cite{llvm_opt_benchmark} (\NumFunctionsShort{}
functions),
where \sys{} identifies the functions each rewrite applies to in isolation
with \MainPrecision{} precision and recall (\cref{subsec:rq2}).
% \cn{
%   What corner cases? It is weird to see corner cases, as this might imply
%   that your evaluation above is not sufficient.
% }
% \xhx{Revised.}
% \cn{
%   It is unclear whether you use the upstream test suites as the corpus,
%   or the query?
% }
% \xhx{I changed the sentence to explicitly say it complements the corpus.}
% \cn{Same queries but just a different corpus?}
% \xhx{
%   yes. some constraints are never satisfied in real-world code,
%   but they are satisfied in the test suite; and vice versa.
% }
% \cn{
%   It is unclear whether the test suites are merged into the large corpus,
%   or used as a separate corpus. It is also unclear whether the same queries
%   are used here.
% }
% \xhx{Revised.}
% \cn{Be careful of using "them". "Them" here is ambiguous.}
% \xhx{Revised.}
\llvm{}'s upstream test suites (\NumTestFiles{} files of hand-written IR)
form a separate corpus that complements \corpusName{}
with forms that real-world code rarely exhibits.
On this corpus,
the same patterns for the \NumPRsWord{} rewrites
achieve \TestPrecision{} precision and \TestRecall{} recall,
with only \TestFN{} false negative on an unsupported vector constant predicate.
The index leaves a median of \MedFuncSurvivePct{} of the functions per pattern,
so a query takes a median of \IndexedSecMedian{} seconds,
\MedIndexSpeedup{} faster than a full scan.
\sys{} is already in use:
in an \llvm{} issue (link withheld for review),
a contributor used it to assess how often a proposed rewrite applies.

\myparagraph{Contributions}
This paper makes the following contributions.
\begin{itemize}[leftmargin=*,nosep]
  \item \textbf{A \dsl{} for both \src{} and \pre{}} (\cref{sec:language}):
  \llvmir{} with holes and constraints that mirror \instcombine{}'s
  matchers and analysis APIs familiar to \llvm{} developers.
  % \item \textbf{A formal semantics} (\cref{sec:semantics})
  % of when a core pattern matches a function,
  % with a proof that lowering the surface language preserves matches.
  % \cn{Is this a key contribution?}
  % \xhx{
  %   Removed.
  %   Maybe not. I just (wrongly) assumed a PLDI paper should add this...
  %   I've checked other PLDI papers, and they usually don't list formal semantics as a contribution.
  % }
  % \cn{
  %   If you think it is interesting, you can append it here under the DSL contribution.
  %   I think it should be fine, but it is an overkill to list it as a separate contribution.
  % }
  % \xhx{not very interesting TBH...}
  \item \textbf{Sound pruning by abstract interpretation}
  (\cref{sec:abstract-interp}).
  The index prunes functions by both structure and preconditions
  without loading their IR.
  This pruning is sound and effective:
  we prove that it never prunes a function that contains a match,
  and preconditions further reduce the candidate functions per pattern
  by a median of \MedNoConstrFactor{} (\cref{subsec:rq3}).
  \item \textbf{An implementation} (\cref{sec:impl}):
  a WebAssembly app that searches the published corpus index in the browser,
  so developers need no local corpus or build.
  A contributor has already used it to assess a proposed \llvm{} rewrite.
  \item \textbf{An evaluation} (\cref{sec:eval}) on \NumPRsWord{} real rewrites
  and \NumFunctionsShort{} functions, which shows that \sys{} is accurate and fast.
  \sys{} patterns match exactly where all \NumExactPRsWord{} rewrites apply
  in isolation, with \MainFP{} false positives against \DaggrepFP{} for \daggrep{}
  at equal recall.
  Queries on \NumThreads{} threads take a median of \IndexedSecMedian{} seconds,
  \MedIndexSpeedup{} faster than a full scan,
  against \DaggrepSecMedian{} seconds for \daggrep{}.
\end{itemize}

\section{Background}
\label{sec:background}

This section reviews the \llvm{} concepts underlying \sys{}:
\llvmir{}, missed peephole optimizations, C++ rewrites in \instcombine{},
and the value-tracking analyses used to check preconditions.

\myparagraph{\llvmir{}}
\llvmir{}~\cite{lattner_llvm_2004} uses static single-assignment form:
each instruction defines at most one value,
and each defined value has a name starting with \code{\%} (\eg{}, \code{\%udiv1}).
An instruction's \emph{opcode} specifies its operation
(\eg{}, \code{add} or \code{udiv}),
and its operands are values defined by other instructions,
function arguments, or constants.
Types include integers of any bit width, such as \code{i32},
and vectors of them, such as \code{<4 x i32>}.
Flags on an instruction state extra properties,
and the instruction's result becomes poison
if a property does not hold.\footnote{%
A poison value is a deferred form of undefined behavior:
it propagates through the instructions that use it,
and it is undefined behavior only if it reaches an operation with a side effect,
such as a branch condition or a store~\cite{llvm_langref_poison}.}
For example, \code{add nsw} yields poison on signed overflow,
and \code{or disjoint} yields poison if its operands share any set bits.
The \emph{uses} of a value are the operand slots that hold the value;
for example, \code{mul \%x, \%x} is two uses of \code{\%x}.

\begin{figure}[t]
\centering
\begin{subfigure}[t]{0.46\linewidth}
\vspace{0pt}
\begin{lstlisting}[style=llvmstyle,escapechar={},basicstyle=\scriptsize\ttfamily,numbers=none,xleftmargin=0pt,aboveskip=0pt,belowskip=0pt]
define i8 @src(i8 %a, i8 %b, i8 %c) {
  %b.small = icmp ult i8 %b, 15
  call void @llvm.assume(i1 %b.small)
  %c.small = icmp ult i8 %c, 15
  call void @llvm.assume(i1 %c.small)
  %udiv1 = udiv i8 %a, %b
  %udiv2 = udiv i8 %udiv1, %c
  ret i8 %udiv2
}

define i8 @tgt(i8 %a, i8 %b, i8 %c) {
  %m = mul nuw i8 %b, %c
  %udiv = udiv i8 %a, %m
  ret i8 %udiv
}
\end{lstlisting}
\caption{Source and Target IR.}
\label{subfig:rewrite-ir}
\end{subfigure}%
\hfill
\begin{subfigure}[t]{0.50\linewidth}
\vspace{0pt}
\begin{lstlisting}[style=cppstyle,basicstyle=\scriptsize\ttfamily,numbers=left,firstnumber=1,numberstyle=\tiny\color{black!50},numbersep=3pt,xleftmargin=1em,aboveskip=0pt,belowskip=0pt]
Instruction *visitUDiv(BinaryOperator &I) {
  Value *Op0 = I.getOperand(0),
        *Op1 = I.getOperand(1);
  ...
  Value *X, *Y;
  if (match(Op0, @\label{ln:cpp-match-start}@
        m_OneUse(@\label{ln:cpp-oneuse}@
          m_UDiv(m_Value(X), m_Value(Y))
        )) @\label{ln:cpp-match-end}@
      && willNotOverflowUnsignedMul(Y, Op1, I)) {@\label{ln:cpp-overflow}@
    Value *YZ = Builder.CreateNUWMul(Y, Op1);@\label{ln:cpp-build-start}@
    return BinaryOperator::CreateUDiv(X, YZ);@\label{ln:cpp-build-end}@
  }
  ...
}
\end{lstlisting}
\caption{C++ Rewrite in \instcombine{}.}
\label{subfig:rewrite-cpp}
\end{subfigure}
\caption{
  A Concrete IR Example illustrating a missed peephole optimization
  (\subref{subfig:rewrite-ir}),
  and the Rewrite implemented in \instcombine{} in \llvmPR{214541}
  realizing this optimization (\subref{subfig:rewrite-cpp}).
It transforms nested unsigned divisions
$(a \udivop b) \udivop c$ into $a \udivop (b \mulop c)$
when the inner division has one use and $b \mulop c$ cannot overflow.
The assumptions in \texttt{src} ensure that the product does not overflow.}
\Description{Two panels: source and target LLVM IR on the left,
and C++ code from InstCombine's visitUDiv on the right.
The source assumes both divisors are below 15 and performs two unsigned divisions.
The target multiplies the divisors and performs one unsigned division.
The C++ code matches Op0 against m_OneUse(m_UDiv(m_Value(X), m_Value(Y))),
checks willNotOverflowUnsignedMul(Y, Op1, I),
and then builds X udiv (Y mul Op1).}
\label{fig:cpp-example}
\end{figure}
\myparagraph{Missed Peephole Optimizations}
The community often receives issues with
% Developers often report
concrete instances of missed peephole optimizations,
such as \cref{subfig:rewrite-ir} from \llvm{} \llvmIssue{132908}.
The \code{src} function computes $(a \udivop b) \udivop c$ with two unsigned divisions;
because the \code{llvm.assume} calls bound both divisors below 15,
their product ($\le 196$) cannot overflow an \code{i8},
allowing \code{tgt} to compute $a \udivop (b \mulop c)$ instead.
\instcombine{} is the primary peephole optimization pass in \llvm{}.
To handle such optimizations in \instcombine{},
developers generalize the concrete example into a rewrite
by abstracting over the \code{i8} type and replacing the specific
\code{llvm.assume} bounds with a general check that $b \mulop c$ cannot overflow.

\myparagraph{Rewrites in \instcombine{}}
\Cref{subfig:rewrite-cpp} shows the C++ rewrite that resolves this issue in
pull request (PR) \llvmPRNum{214541}.
\instcombine{} visits each instruction and calls the handler for its opcode,
such as \cpp{visitUDiv} for the outer \code{udiv} (\code{\%udiv2}).
Lines~\ref{ln:cpp-match-start}--\ref{ln:cpp-match-end} match \src{}
with \llvm{}'s \patternmatch{} matcher library~\cite{llvm_patternmatch},
whose matchers compose recursively by taking sub-matchers as arguments:
\cpp{m_UDiv(m_Value(X), m_Value(Y))} matches \cpp{Op0} (\code{\%udiv1}) as a \code{udiv}
and uses \cpp{m_Value} to bind \cpp{Op0}'s operands to \cpp{X} (\code{\%a}) and \cpp{Y} (\code{\%b}).
Wrapping \cpp{m_UDiv} in \cpp{m_OneUse} (line~\ref{ln:cpp-oneuse}) requires the
inner division~\cpp{Op0} to have a single use (as the dividend of~\cpp{I}).
If \cpp{Op0} has more than one use, \cpp{Op0} could not be removed,
because the rewrite would add a multiplication without eliminating a division.
Finally, line~\ref{ln:cpp-overflow} completes \pre{} by checking that
$\cpp{Y} * \cpp{Op1}$ (\code{\%b} and \code{\%c}) cannot overflow,
and lines~\ref{ln:cpp-build-start}--\ref{ln:cpp-build-end} build the \tgt{}.

\myparagraph{Value Tracking}
Preconditions that \patternmatch{} cannot check,
such as ruling out overflow or requiring a nonzero operand,
rely on dataflow facts.
\instcombine{} obtains these facts from \llvm{}'s \valuetracking{} analyses,
which infer known bits and value bounds
from defining instructions and branch conditions.
Because a fact may hold only at certain points in a function,
each \valuetracking{} query takes a \emph{context instruction}
to check whether the fact holds there.
% \xt{
%   On line~\ref{ln:cpp-overflow} of \cref{subfig:rewrite-cpp},
%   \cpp{willNotOverflowUnsignedMul(Y,Op1,I)} checks that
%   $\cpp{Y} * \cpp{Op1}$ will not overflow at context~\cpp{I}.
%   For the IR in \cref{subfig:rewrite-ir},
%   this asks whether $\code{\%b} * \code{\%c}$ can overflow at \code{\%r}.
%   Both assumptions dominate \code{\%r},
%   allowing \valuetracking{} to establish the bounds on the divisors
%   and prove that their product cannot overflow.
% }
% \xhx{
%   I didn't adopt your draft here because I think using llvm.assume is
%   not a good or real example.
% }
% \xt{
%   You are right,
%   it's not real-world IR but just
%   to represent the precndition.
%   Closing the comments.
% }
On line~\ref{ln:cpp-overflow} of \cref{subfig:rewrite-cpp},
\cpp{willNotOverflowUnsignedMul(Y,Op1,I)} uses \valuetracking{}
to check that $\cpp{Y}*\cpp{Op1}$ will not overflow
at context instruction~\cpp{I}.
For example, if \cpp{Y} is zero-extended from 16 to 32 bits
and \cpp{Op1} is 7,
\valuetracking{} knows from \cpp{Y}'s defining instruction
that its upper 16 bits are zero, so $\cpp{Y} < 2^{16}$;
their 32-bit unsigned product is therefore below $7 \cdot 2^{16} < 2^{32}$
and cannot overflow.

\section{Overview}
\label{sec:overview}

% \xt{
%   I found that we start with the C++ code (Figure 2)
%   and the \sys{} pattern,
%   without an initial concrete example of the IR that matches.
%   First, this give an impression that the pattern
%   is derived from the C++ code,
%   and has been implemented in LLVM.
%   Also, I remember we claimed somewhere that
%   ``the developer can start from a concrete example
%   and then iteratively generalize it into a pattern''.
%   We currently fail to show this.
% }

This section illustrates \sys{} using the missed optimization in
\cref{subfig:rewrite-ir}.
A developer first expresses the optimization's applicability conditions
as a \sys{} pattern.
The developer then evaluates the pattern over a large corpus accurately
and efficiently before implementing the optimization in C++ in \llvm{}.

% This section illustrates \sys{} with the rewrite
% of \llvmPR{214541} (\cref{fig:cpp-example}).

\myparagraph{Writing a \sys{} Pattern}
Starting from the function \code{@src} in \cref{subfig:rewrite-ir},
a developer generalizes the concrete IR into a \emph{surface pattern}
(\cref{fig:overview-example}), written in \sys{}'s \dsl{}
(\cref{sec:language}) to specify both \src{} and \pre{}.
To express \src{}, \sys{} uses \emph{\metavars{}} (names starting with
\code{?}) as holes.
The operands \code{\%a}, \code{\%b}, and \code{\%c} become \code{?x},
\code{?y}, and \code{?z}, each matching any value (like \cpp{m_Value}).
Likewise, the type \code{i8} becomes \code{i?bits},
matching an integer type of any width and binding that width to \code{?bits}.
Names starting with \code{\%}
% \cn{
%   Why not use \%q and \%r to minimize the mental burden?
% }
(\code{\%udiv1} and \code{\%udiv2}) are \emph{local names}
for the pattern's instructions.
Subsequent instructions and constraints refer to instructions
by these local names (\eg{}, \code{\%udiv2} takes \code{\%udiv1} as its
dividend).
To express \pre{}, \sys{} provides \emph{\metaflags{}} and \emph{constraints}.
A \metaflag{}, written after a colon, refines an instruction or operand
inline.
For example, \code{:s} on \code{\%udiv1} requires the inner division
to have a single use
(like \cpp{m_OneUse} on line~\ref{ln:cpp-oneuse} of \cref{subfig:rewrite-cpp}).
A line starting with \code{;;} adds a constraint.
Rather than matching the specific \code{llvm.assume} calls in \code{@src},
\code{noOverflow("umul", ?y, ?z)} queries \valuetracking{} to check that
the unsigned product $\code{?y} * \code{?z}$ cannot overflow
(like \cpp{willNotOverflowUnsignedMul} on line~\ref{ln:cpp-overflow} of
\cref{subfig:rewrite-cpp}).

\newsavebox{\mpirbox}
\begin{figure}[t]
  \centering
  % Keep circled numbers at the text size, independent of the stage box dimensions.
  \renewcommand*{\circled}[1]{%
    \tikz[baseline=(char.base)]{%
      \node[shape=circle, draw, minimum size=0pt, text width=,
        inner sep=0.1ex, font={}] (char) {#1};}}
\begin{lrbox}{\mpirbox}\begin{minipage}{38mm}
\begin{lstlisting}[style=mpirstyle,basicstyle=\scriptsize\ttfamily,xleftmargin=0pt,aboveskip=0pt,belowskip=0pt]
%udiv1 = udiv@\hlc{\texttt{:s}}@ i?bits ?x, ?y
%udiv2 = udiv i?bits %udiv1, ?z
;; noOverflow("umul", ?y, ?z)
\end{lstlisting}
\end{minipage}\end{lrbox}%
  \begin{tikzpicture}[
      stage/.style={draw, rounded corners, align=center, font=\footnotesize,
      minimum height=7mm, text width=41mm, inner sep=1.5pt,
      fill=BlindColorTolOne!12},
      io/.style={draw=black!50, densely dotted, rounded corners=1.5pt,
      align=left, font=\scriptsize, inner sep=2pt},
      arr/.style={-{Stealth[length=1.6mm]}},
    ]
    \matrix[column sep=5mm, row sep=2mm, ampersand replacement=\&,
      nodes={anchor=center}] {
      % Stage 1.
      \node[io, text width=38mm] (pat)
        {\textbf{surface pattern} (\sys{}):\\[1pt]\usebox{\mpirbox}}; \&
      \node[stage] (low)
        {\textbf{\StageLower{}: Lowering}\\{\scriptsize into \variants{}}}; \&
      \node[io, text width=42mm] (out1)
        {\textbf{\variant{}}:\\
         \fc{\%udiv1 = udiv i?bits ?x, ?y}\\
         \fc{\%udiv2 = udiv i?bits \%udiv1, ?z}\\
         \hlc{\fc{;; hasOneUse(\%udiv1)}}\\
         \fc{;; noOverflow("umul", ?y, ?z)}}; \\
      % Stage 2.
      \node[io, text width=38mm] (in2)
        {\textbf{posting lists}:\\
         \fc{udiv}$\to$\fc{udiv}: \texttt{@f}, \texttt{@g}, \texttt{@h}\\
         \textcolor{black!50}{\texttt{udiv}$\to$\texttt{zext}: \texttt{@f}, \texttt{@k}\quad\ldots}}; \&
      \node[stage] (look)
        {\textbf{\StageLookup{}: Index Lookup}\\{\scriptsize inverted index of posting lists}}; \&
      \node[io, text width=42mm] (out2)
        {\textbf{candidates}: \texttt{@f}, \texttt{@g}, \texttt{@h}}; \\
      % Stage 3.
      \node[io, text width=38mm] (in3)
        {\textbf{\features{}}:\\
         \texttt{@f}, \texttt{@g}: a single-use \fc{udiv}\\
         \texttt{@h}: no single-use \fc{udiv}}; \&
      \node[stage] (filt)
        {\textbf{\StageFilter{}: Feature Filtering}\\{\scriptsize constraints on \features{}}}; \&
      \node[io, text width=42mm] (out3)
        {\textbf{candidates}: \texttt{@f}, \texttt{@g}\\
         \texttt{@h}: pruned, \fc{hasOneUse(\%udiv1)} is false}; \\
      % Stage 4.
      \node[io, text width=38mm] (in4)
        {\textbf{function IR}:\\
         \texttt{@f}: \fc{\%b = zext i16 \%s to i32}\\
         \hphantom{\texttt{@f}: }\fc{\%q = udiv i32 \%a, \%b}\\
         \hphantom{\texttt{@f}: }\fc{\%r = udiv i32 \%q, 7}\\[1pt]
         \texttt{@g}: \fc{\%t = udiv i32 \%p, \%m}\\
         \hphantom{\texttt{@g}: }\fc{\%u = udiv i32 \%t, \%n}}; \&
      \node[stage] (struct)
        {\textbf{\StageMatch{}: Structural Matching}\\{\scriptsize pattern on IR}}; \&
      \node[io, text width=42mm] (out4)
        {\textbf{\instantiations{}}:\\
         \begin{tabular}{@{}l@{\enspace}r@{\,$\mapsto$\,}l@{\quad}r@{\,$\mapsto$\,}l@{}}
           \texttt{@f}: & \fc{\%udiv1} & \fc{\%q} & \fc{\%udiv2} & \fc{\%r} \\
            & \fc{?x} & \fc{\%a} & \fc{?y} & \fc{\%b} \\
            & \fc{?z} & \fc{7} & \fc{?bits} & 32 \\[1pt]
           \texttt{@g}: & \multicolumn{4}{@{}l@{}}{likewise,
             \fc{?y}\,$\mapsto$\,\fc{\%m}, \fc{?z}\,$\mapsto$\,\fc{\%n}} \\
         \end{tabular}}; \\
      % Stage 5.
      \& \node[stage] (cons)
        {\textbf{\StageEval{}: Constraint Evaluation}\\{\scriptsize constraints on IR}}; \&
      \node[io, text width=42mm] (out5)
        {\textbf{matches}: \texttt{@f} at \fc{\%r}\\
         \texttt{@g}: rejected, \fc{noOverflow} is false}; \\
    };
    \draw[arr] (low) -- (look);
    \draw[arr] (look) -- (filt);
    \draw[arr] (filt) -- (struct);
    \draw[arr] (struct) -- (cons);
    \foreach \i/\s in {pat/low, in2/look, in3/filt, in4/struct}
      \draw[arr, shorten <=0.5mm] (\i.east |- \s) -- (\s);
    \draw[arr, shorten <=0.5mm] (in4.south) |- (cons.west);
    \foreach \s/\o in {low/out1, look/out2, filt/out3, struct/out4, cons/out5}
      \draw[arr, shorten >=0.5mm] (\s) -- (\o.west |- \s);
  \end{tikzpicture}
  \caption{The Running Example from the Rewrite of \llvmPR{214541}.
  Stages run from top to bottom; each reads the input shown on its left and produces the result shown on its right.
  Yellow highlighting marks how lowering converts \texttt{:s} into a constraint.
  An arrow \texttt{udiv}$\to$\texttt{udiv} means that a \texttt{udiv} uses another
  \texttt{udiv}.}
  \Description{A pipeline diagram for the running example. The MetaphIR pattern is the input of lowering, shown on the left. Five stages run top to bottom: lowering, index lookup, feature filtering, structural matching, and constraint evaluation. Each stage shows its result on the right; index lookup reads a posting list, feature filtering the features of the remaining functions, and structural matching the function IR, also shown on the left. Only function f matches.}
  \label{fig:overview-example}
\end{figure}

\myparagraph{Querying the Corpus}
To see how widely this optimization applies,
a developer runs the pattern as a \emph{query} over \corpusName{}
(\NumFunctionsShort{} functions from \NumProjects{} real-world projects,
\cref{sec:eval}).
For this pattern, \sys{} reports matches in \RunPatternBenchMatches{} functions
in \IndexedSec{} seconds,
identifying the exact functions where the C++ rewrite applies.
\Cref{fig:overview-example} illustrates the \NumStagesWord{} stages that
execute this query, starting by lowering the surface pattern into simpler
\emph{\variants{}} and running them on a small example corpus in which
only \code{@f} matches.

\itemParagraph{\StageLower{}: Lowering}
\sys{} separates the surface language from a minimal \emph{core language}
(\cref{sec:language,sec:semantics}) for extensibility and maintainability:
users can define new abstractions in the surface language,
while downstream indexing, matching, and soundness proofs
(\cref{subsec:evaluator}) need only handle a small set of core primitives.
Lowering expands a surface pattern into a set of \variants{},
each containing only primitive instructions with \metavars{} and constraints.
In \cref{fig:overview-example},
lowering converts the \metaflag{} \code{:s} into the constraint
\code{hasOneUse(\%udiv1)} and produces only one \variant{}.
A surface pattern can lower into multiple \variants{}
when it uses a \emph{\composite{}} with alternatives,
such as \code{addLike}, which matches either \code{add} or \code{or disjoint}.

\itemParagraph{\StageLookup{}: Index Lookup}
Reading IR is the bottleneck of query execution:
the corpus contains \CorpusSizeGB{}~GB of bitcode,
and each module must be loaded and parsed before matching,
even though most functions do not match.
To prune non-matching functions without reading their IR,
\sys{} builds a corpus index in two parts, used by \StageLookup{} and \StageFilter{}, respectively.
The first part is an inverted index of \emph{posting lists}.
Each posting list maps a structural \feature{}
directly to the functions that exhibit it.
Examples of structural \features{} are the opcodes of a function's instructions
and the opcode pairs of instructions where one uses the other;
\cref{subsec:features} lists all of them.
A pattern usually requires several structural \features{},
and \sys{} intersects their posting lists.
In \cref{fig:overview-example},
the posting list for the opcode pair \code{udiv}$\to$\code{udiv}
(one \code{udiv} using another \code{udiv})
contains \code{@f}, \code{@g}, and \code{@h}.
These form the initial \emph{candidates}.

\itemParagraph{\StageFilter{}: Feature Filtering}
The second part of the index stores each function's \emph{\features{}},
which include its structural \features{}
and the \emph{\facts{}} that hold for the values in each operand slot.
\Features{} are compact.
The whole index takes \IndexSizeGB{}~GB against \CorpusSizeGB{}~GB of bitcode,
and decoding a function's \features{} requires no IR parsing.
\sys{} decodes the \features{} of only the candidates
from \StageLookup{}
and evaluates the query constraints on them.
This evaluation is an abstract interpretation~\cite{cousot_abstract_1977}
of the constraint language.
A constraint evaluates to true, false, or unknown,
and \sys{} prunes a function only when a constraint is false.
For \code{@h}, the \features{} show that no \code{udiv} has a single use,
so \code{hasOneUse(\%udiv1)} is false and \code{@h} is pruned,
leaving \code{@f} and \code{@g}.
\Cref{sec:abstract-interp} proves this abstract interpretation sound,
\ie{}, it never prunes a function that has a match.

\itemParagraph{\StageMatch{}: Structural Matching}
Only at this stage does \sys{} read any IR.
On the full corpus, \RunPatternFuncs{} of \NumFunctionsShort{} functions
remain as candidates, in \RunPatternMods{} modules,
so \sys{} loads only these modules.
\sys{} searches each candidate for \emph{\instantiations{}}:
maps from the pattern's instructions and \metavars{} to IR values
that respect opcodes, types, flags, and operand edges.
In \cref{fig:overview-example},
\code{@f} computes \code{\%q = udiv i32 \%a, \%b}
and then
\code{\%r = udiv i32 \%q, 7}.
% \lch{The following sentence is a bit crowded.}
% \xhx{Revised.}
The \instantiation{} for \code{@f} maps the two divisions of the pattern
to \code{\%q} and \code{\%r},
and each \metavar{} to the matching operand or width,
\eg{}, \code{?z} to \code{7}.
Function \code{@g} has the same shape and also yields one \instantiation{}.

\itemParagraph{\StageEval{}: Constraint Evaluation}
Finally, \sys{} evaluates each constraint on each \instantiation{}.
For \code{@f},
\code{hasOneUse(\%q)} checks the use list of \code{\%q},
and \code{noOverflow} queries \valuetracking{} at context \code{\%r}:
because \code{\%b} is zero-extended from 16 bits,
$\code{\%b} * 7$ cannot overflow (\cref{sec:background}) and \code{@f} matches.
In \code{@g}, nothing bounds \code{\%m} and \code{\%n},
so \code{noOverflow} is false and \code{@g} is rejected.
An \instantiation{} that passes every constraint is a match.

% \myparagraph{Comparison with \daggrep{}}
% \daggrep{} cannot state the single-use and no-overflow conditions,
% so it also reports division chains where the fold does not apply,
% and it scans every module on each query (\cref{subsec:rq2,subsec:rq3}).

\section{The \sys{} Language}
\label{sec:language}
We design the \sys{} \dsl{} to be immediately familiar to \llvm{} developers:
patterns use \llvmir{}-like syntax to mirror the instructions they match;
built-in \metaflags{} and constraints borrow the names and semantics of
\patternmatch{} matchers, \valuetracking{} analyses, and \llvm{} utility APIs.
% so a developer who knows \llvm{} can often predict the \sys{} spelling.
\Cref{tbl:patternmatch} lists common \patternmatch{} idioms and their \sys{}
counterparts.

For extensibility and maintainability,
the \sys{} \dsl{} (the \emph{surface language}) layers three abstractions
(\cref{subsec:lang-abstraction}) over a small \emph{core language}
(\cref{subsec:lang-core}):
\sys{}'s standard library and users can define new abstractions in the surface language,
while indexing, matching, and soundness proofs need only handle the core primitives.
Lowering (\cref{subsec:lowering}) converts \emph{surface patterns} into
\emph{\variants{}} in the core language,
which we formalize in \cref{sec:semantics}.

\begin{table}[t]
  \footnotesize
  \centering
  \caption{\patternmatch{} Idioms and Their \sys{} Counterparts.}
  \label{tbl:patternmatch}
  \begin{tabular}{@{}lll@{}}
    \toprule
    \textbf{\patternmatch{} (C++)} & \textbf{\sys{}} & \textbf{Meaning} \\
    \midrule
    \hcpp{m_Value(X)}, \hcpp{m_Deferred(X)} & \hcode{?x}, its second occurrence & any value; the same value \\
    \hcpp{m_Specific(V)} & \hcode{\%v} & the value of a pattern instruction \\
    \hcpp{m_c_Add(A, B)} & \hcode{add:c ?ty A, B} & commutative matching \\
    \hcpp{m_OneUse(P)} & \hcode{P:s} & a single use \\
    \hcpp{m_ZExtOrSelf(P)} & \hcode{zext:o ...} & the instruction or its operand \\
    \hcpp{m_NSWAdd}, ``no \hcode{nsw}'', a saved flag & \hcode{add nsw}, \hcode{add !nsw}, \hcode{add nsw:?f} & flag present, absent, captured \\
    \hcpp{m_APInt(C)}, \hcpp{m_Power2(C)} & \hcode{?c:apInt}, \hcode{?c:power2} & constant classes \\
    \hcpp{m_AddLike}, \hcpp{m_Shift}, \hcpp{m_Signum} & \hcode{addLike}, \hcode{shift}, \hcode{signum} & \composites{} \\
    \hcpp{isKnownNonZero(X, Q)} & \hcode{isKnownNonZero(?x)} & value tracking at the context \\
    \hcpp{C->popcount() == *S} & \hcode{popcount(?c) == ?s} & a constraint \\
    \bottomrule
  \end{tabular}
\end{table}

\subsection{The Core Language}
\label{subsec:lang-core}

A \variant{} is a sequence of primitive instructions in \llvmir{} syntax,
followed or interleaved by Boolean constraints.

\subsubsection{Patterns Are IR with Holes}
\label{subsec:lang-ir}

The core language covers \NumInstMatchers{} \llvm{} instructions,
including arithmetic, bitwise and vector operations, casts, compares,
\code{select}, \code{phi}, \code{call}, \code{br}, and memory accesses.
Anything concrete in an instruction must match exactly,
and a \metavar{} \code{?name} is a hole.
Each \metavar{} has a \emph{sort}, inferred from where it appears
(\cref{tbl:sorts}).
All occurrences of a \metavar{} must bind the same thing.
% \xt{
%   Why not generalize to multiple times?
% }
% \xhx{revised.}

\begin{table}[t]
\centering
\caption{
Sorts of \TMetavars{}
(sorts may have subsorts and thus overlap, \eg{}, $\sInst \subsort \sVal$, \cref{subsec:core-sorts}).
% \lch{Val and Inst are both able to match an instruction. Not sure whether it might be better to briefly show their difference? }
% \xhx{Added a note on subsorts.}
}
\label{tbl:sorts}
\footnotesize
\begin{tabular}{@{}lp{0.3\linewidth}>{\raggedright\arraybackslash}p{0.5\linewidth}@{}}
  \toprule
  \textbf{Sort} & \textbf{Example} & \textbf{Binds} \\
  \midrule
  \sVal & \hcode{add i8}\hsp\hmv{?x}\hcode{,}\hsp\hmv{?y} & an instruction, argument, constant, or global \\
  \sNat & \hcode{add i}\hmv{?bits}\hsp\hcode{...}, \hcode{<}\hmv{?n}\hsp\hcode{x i32>} & a bit width, count, or index \\
  \sType & \hcode{add}\hsp\hmv{?ty}\hsp\hcode{...} & an \llvm{} type, such as \code{i32} \\
  \sPred & \hcode{icmp}\hsp\hmv{?pred}\hsp\hcode{i32 ...} & a comparison predicate \\
  \sBool & \hcode{add nsw:}\hmv{?f}\hsp\hcode{...} & whether a flag is present \\
  \sInst & \hcode{add:}\hmv{?instr}\hsp\hcode{...} & the matched instruction \\
  \sBB & \hcode{br label}\hsp\hmv{?bb} & a basic block \\
  \sFn & \hcode{call}\hsp\hmv{?f}\hcode{(...)} & a callee \\
  \sIntr & \hcode{call}\hsp\texttt{@llvm.}\hmv{?id}\hcode{(...)} & an intrinsic \\
  \kw{Type[]}, \kw{Val[]} & \hcode{call i32}\hsp\texttt{@f(...}\hmv{?tys}\hsp\hmv{?vals}\hcode{)} & the \llvm{} types and the values of the remaining operands \\
  \bottomrule
\end{tabular}
\end{table}

\myparagraph{Flags}
\llvm{} instructions carry flags, such as \code{nsw} on \code{add}
or \code{exact} on \code{udiv}, which often make the instruction
yield a poison value when the flag's specified property does not hold.
A pattern specifies each flag in one of four ways:
% \xt{
%   I suggest consistently using
%   ``\code{?x, ?y}'' or ``\code{...}''
%   in all four cases.
% }
% \xhx{Make sense. Revised.}
\begin{itemize}[leftmargin=*,nosep]
  \item omit it (\code{add i32 ?x, ?y}): the flag may or may not be present;
  \item write it (\code{add nsw i32 ?x, ?y}): the flag must be present;
  \item negate it (\code{add !nsw i32 ?x, ?y}): the flag must be absent;
  \item capture it (\code{add nsw:?f i32 ?x, ?y}):
  the flag may or may not be present,
  and \code{?f} binds to \code{true} if it is present and \code{false} otherwise.
  % \xt{
  %   This is a bit vague.
  %   What does ``either way'' mean?
  % }
  % \xhx{Revised.}
\end{itemize}
% These four cases correspond to the ways \patternmatch{} code inspects flags:
% \cpp{m_Add} ignores them, \cpp{m_NSWAdd} requires one,
% and an explicit \cpp{hasNoSignedWrap()} test either rejects the flag
% or saves its value for later use.

\myparagraph{Structural \Metaflags{}}
% \xt{
%   I think this introduces confusion.
%   It's unclear what the difference is between a structural
%   metaflag and other metaflags.
%   You claimed
%   ``
%   Each \variant{} is a list of instructions with \metavars{} and a list of constraints,
%   where every \metaflag{} is lowered into a constraint.
%   ''
%   in the ``Stage 1: Lowering.'' parapraph,
%   but here, as well as the beginning of section 4.3,
%   you define structural metaflags as part of the core language.
% }
% \xhx{Clarified here and removed that claim in Stage 1.}
Two built-in \emph{structural \metaflags{}} (\code{:c} and \code{:o}) are part
of the core language and change how an instruction matches
(unlike user- and library-defined \metaflags{} such as \code{:s}, which lower
into constraints, \cref{subsubsec:lang-metaflags}).
The \metaflag{} \code{:c} tries the operands in swapped order\footnote{On a comparison instruction (\eg{}, \code{icmp}),
swapped operands also swap the predicate (\eg{}, \code{slt} becomes \code{sge}).},
like the \cpp{m_c_} matchers in \patternmatch{}.
The \metaflag{} \code{:o} (``or self'') makes the instruction optional:
\code{zext:o i8 \%t to i32} matches either a \code{zext} of \code{\%t}
or \code{\%t} itself, like \cpp{m_ZExtOrSelf}.

\myparagraph{Blocks}
By default (or after \code{anyblock:}), instructions may lie in any basic block.
A label \code{?bb:} restricts subsequent instructions to a single basic block
and binds that block to \code{?bb}.

\subsubsection{Constraints}
\label{subsec:lang-constraints}

A line \code{;; e} adds the constraint \code{e},
a Boolean expression over the \metavars{}.
The constraint language has integer, Boolean,
and string literals;
arithmetic, bitwise,
and comparison operators;
and \NumOperators{} built-in operators.
The built-in operators cover four groups:
\begin{itemize}[leftmargin=*,nosep]
  \item \emph{Constant classes}, such as \code{isPower2}, \code{isLowBitMask},
  and \code{isSignMask}.
  Each has the semantics of the corresponding \patternmatch{} matcher
  (\code{isPower2} corresponds to \cpp{m_Power2}).
  % including its treatment of vector splats and poison lanes.
  \item \emph{Inspection} of what a \metavar{} bound.
  The constraints \code{opcode}, \code{predicate}, and \code{isa} tell which instruction,
  predicate, or class of value was matched.
  % for example when a \composite{} accepts several opcodes;
  The constraint \code{bitWidth} gives a width of a bound type or value,
  and use counts such as \code{hasOneUse} limit how often a value is used.
  \item \emph{Predicate algebra} on a comparison predicate bound by a
  \metavar{}, as \code{?p} in \code{icmp ?p}.
  The constraint \code{isSigned(?p)} accepts only signed predicates
  such as \code{slt} and \code{sgt},
  and \code{swappedPredicate} checks whether two comparisons are the same
  with their operands swapped.
  These mirror the \cpp{CmpInst} methods \cpp{isSigned} and \cpp{getSwappedPredicate}
  in C++ rewrites.
  \item \emph{Value tracking}:
  \NumValueTrackOps{} queries that call \llvm{}'s \valuetracking{} analyses,
  such as \code{isKnownNonZero}, \code{knownBits}, \code{noOverflow},
  and \code{inUnsignedRange}.
  Each query's \emph{context instruction} (\cref{sec:background}) defaults to
  the instruction written immediately before the constraint (\eg{},
  \code{\%udiv2} for \code{noOverflow} in \cref{fig:overview-example}).
  % or can be passed explicitly as a final argument.
\end{itemize}

% \lch{Personally, I can understand the following paragraphs, but
% their relevance isn't clear to me. I'm not sure why they need
% to be included and it makes the this section not systematic enough.
% Can’t we just keep the text above?}
% \xhx{Blended Context Instruction into the Value tracking bullet and merged the arithmetic/coercion notes into the intro paragraph below.}
In a constraint, a \metavar{} stands for the object it bound;
arithmetic and comparison operators implicitly coerce a bound IR value to its
integer value if it is a constant, and reject the match if coercion fails
(\cref{subsec:core-sorts}).
Arithmetic uses unbounded integers by default (so \eg{} \code{?bits + 1} never
overflows), while fixed-width wrapping operations take the bit width explicitly
(\eg{}, \code{add(128, 128, 8) == 0}).

\subsection{Extending the Core}
\label{subsec:lang-abstraction}
% \lch{I think it might be better to say why we need to
% extend the core in the first paragraph more explicitly.
% Currently, I feel we have put a lot of text in introducing the core language
% but not clearly say why we need to extend it.}
% \xhx{Revised.}
Rather than hardcoding every \patternmatch{} idiom,
% \xt{
%   What's the core evaluator?
% }
% \xhx{Changed to "as built-in operators".}
the core language provides only minimal primitives.
\sys{} extends the core into the surface language with three abstractions:
named constraints, \metaflags{}, and \composites{}.
The standard library \stdlib{} uses them to define the counterparts of
\patternmatch{} (\eg{}, \code{:s} and \code{addLike}),
and users can define their own in the same way.

\subsubsection{Named Constraints and \Metaflags{}}
\label{subsubsec:lang-constraints}
\label{subsubsec:lang-metaflags}

A named constraint is a reusable Boolean expression,
and a \metaflag{} attaches a constraint to the value it follows,
which keeps patterns short and readable.
Below, the named constraint \code{inRange} tests a range,
and the \metaflag{} \code{:between} applies \code{inRange} to the flagged value.
The first parameter of a \metaflag{} (\code{?v})
is the operand or instruction that the \metaflag{} is attached to,
and the remaining parameters are written in parentheses,
so \code{?c:between(5, 10)} requires \code{?c} to lie in $[5, 10]$:

\noindent\begin{minipage}{\linewidth}
\begin{lstlisting}[style=mpirstyle]
constraint inRange(?v, ?lo, ?hi) { sle(?lo, ?v) && sle(?v, ?hi) }
metaflag :between(?v, ?lo, ?hi) { inRange(?v, ?lo, ?hi) }
%r = add i32 ?x, ?c:between(5, 10)
\end{lstlisting}
\end{minipage}

\noindent A \metaflag{} declaration also says where the \metaflag{} may be
attached:
to operands (the default), to instructions (\code{metaflag:i}),
or to both (\code{metaflag:i:o}).
% The standard library builds \NumStdConstraints{} named constraints from the
% built-in operators, such as \code{isAPInt} for \cpp{m_APInt},
% and \NumStdMetaflags{} \metaflags{},
% such as \code{:s} for a single use and \code{:power2} for \cpp{m_Power2}.
\Metaflags{} combine naturally:
a value with several \metaflags{} must satisfy all of them,
\code{:or(...)} accepts a value that satisfies any one of them,
and \code{:not(...)} accepts a value that fails the given \metaflag{}.

\subsubsection{\TComposites{}}
\label{subsubsec:lang-composites}

A \code{match} declaration defines a new instruction-like matcher with one
or more alternative bodies, called \emph{branches}, separated by \code{or}.
The standard library's \code{addLike}, the counterpart of \cpp{m_AddLike},
accepts an \code{add} or an \code{or disjoint}:

\noindent\begin{minipage}{\linewidth}
\begin{lstlisting}[style=mpirstyle]
match:c addLike [nsw, nuw] Binary (?ty ?x, ?y) { %ret = add nsw:?nsw nuw:?nuw ?ty ?x, ?y }
        or [nsw = true, nuw = true] { %ret = or disjoint ?ty ?x, ?y }
\end{lstlisting}
\end{minipage}

\noindent The declaration specifies that \code{addLike} is commutative (\code{match:c})
and supports the flags \code{nsw} and \code{nuw};
the instruction named \code{\%ret} gives the result of each branch.

\myparagraph{Invocation Forms}
The \emph{form} fixes the shape of an invocation.
\code{Binary} requires two operands of one type \code{?ty},
so an invocation is written like a binary operation,
as in \code{addLike nsw i32 \%a, \%b}.
\Cref{tbl:archetypes} lists the four forms.
% Each \code{Nary} operand has its own type,
% which may be fixed (\code{i1}) or shared (\code{?ty} in \code{(i1 ?c, ?ty ?t,
% ?ty ?f)}).

\begin{table}[t]
  \footnotesize
  \centering
  \caption{Invocation Forms of \TComposites{}.}
  \label{tbl:archetypes}
  \begin{tabular}{@{}llll@{}}
    \toprule
    \textbf{Form} & \textbf{Signature} & \textbf{Example Invocation} & \textbf{\stdlib{} Matchers} \\
    \midrule
    \hcode{Unary} & \hcode{(?ty ?x)} & \hcode{neg nsw i32 ?a} & \hcode{neg}, \hcode{signum}, \hcode{fnegLike} \\
    \hcode{Binary} & \hcode{(?ty ?x, ?y)} & \hcode{addLike nsw i32 ?a, ?b} & \hcode{addLike}, \hcode{shift}, \hcode{idiv} \\
    \hcode{Cast} & \hcode{(?ty1 ?x to ?ty2)} & \hcode{zextOrSext i8 ?a to i32} & \hcode{zextOrSext}, \hcode{sextLike}, \hcode{itofp} \\
    \hcode{Nary} & \hcode{(?ty1 ?x1, ?ty2 ?x2, ...)} & \hcode{selectEq i8 ?x, i32 ?a, i32 ?b} & \hcode{selectEq}, \hcode{selectEqZero} \\
    \bottomrule
  \end{tabular}
\end{table}

\myparagraph{Flag Binding}
An invocation may specify each declared flag (\eg{}, \code{nsw} and \code{nuw})
in the same four ways as an instruction:
omitted, required, negated, or captured.
In the above example,
the first branch passes these flags on to its \code{add}.
An \code{or disjoint} has no such flags,
so the second branch fixes both to true,
since a disjoint \code{or} never wraps.
In this way, the flags are kept meaningful for every branch:
an invocation that writes \code{!nsw} or \code{!nuw} drops the second branch,
and an invocation that captures a flag, as in \code{addLike nsw:?f},
binds \code{?f} to true when the second branch matches.

% \myparagraph{Multi-Instruction Bodies}
% A branch may contain several instructions and constraints.
% The body of the standard library's \code{signum},
% the counterpart of \cpp{m_Signum},
% has four instructions and one constraint.

\section{Formal Semantics}
\label{sec:semantics}

This section formalizes the core language of \sys{},
defines when a \variant{} matches an IR function,
and shows that lowering preserves the matches of a surface pattern.

\begin{figure}[t]
\footnotesize
\renewcommand{\arraystretch}{1}
\[
\begin{array}{@{}r@{\;}c@{\;}l@{\quad}l@{\quad}l@{}}
  \multicolumn{3}{@{}l}{\text{\textbf{Syntax}}} & \text{\textbf{Concept}} & \text{\textbf{Concrete Example}} \\
  \mv & \in & \MVars & \text{\metavars{}} & \text{\footnotesize\hcode{?x}} \\
  \lv & \in & \LNames & \text{local names} & \text{\footnotesize\hcode{\%udiv1}} \\
  k & \in & \Consts & \text{literals and constants} & \text{\footnotesize\hcode{0},\enspace\hcode{\"umul\"},\enspace\hcode{poison}} \\
  f & \in & \Ops & \text{built-in operators} & \text{\footnotesize\hcode{popcount},\enspace\hcode{noOverflow}} \\
  \eta & ::= & n \mid \mv \quad (n \in \mathbb{N}) & \text{sizes} & \text{\footnotesize\hcode{32},\enspace\hcode{?bits},\enspace\hcode{?n}} \\
  \tau & ::= & \kw{i}\,\eta \mid \kw{float} \mid \kw{ptr} \mid \langle \eta \times \tau \rangle \mid \mv \mid \cdots & \text{type patterns} & \text{\footnotesize\hcode{i?bits},\enspace\hcode{<?n x i8>}} \\
  o & ::= & \mv \mid \lv \mid k & \text{operands} & \text{\footnotesize\hcode{?z},\enspace\hcode{\%udiv1},\enspace\hcode{0}} \\
  \flagspec & \in & \FlagSet \rightharpoonup \{+, -\} \cup \MVars & \text{flag specifications} & \text{\footnotesize\hcode{nsw !nuw},\enspace\hcode{nsw:?f}} \\
  \mks & \subseteq & \{\mkc\} \cup \{\mko{o} \mid o \text{ is an operand}\} & \text{structural marks} & \text{\footnotesize\hcode{:c},\enspace\hcode{:o}},\enspace\hcode{:o(2)} \\
  \keyspec & \in & \KeySet \rightharpoonup \Consts \cup \MVars & \text{keyed arguments} & \text{\footnotesize\hcode{icmp}~\hlc{\hcode{?p}} \hcode{...},\enspace\hcode{load ...}~\hlc{\hcode{align 4}}} \\
  I & ::= & \lv = \op^{\mks}_{\flagspec,\keyspec}\,[\tau_0]\;(\tau_1\,o_1, \ldots, \tau_n\,o_n)\;[\kwas\;\mvi]\;[\kwin\;\mvbb] & \text{instruction nodes} & \text{\footnotesize\hcode{\%r = udiv i8 \%t, ?z}} \\
  e & ::= & k \mid \mv \mid \lv \mid f(e_1, \ldots, e_n) \mid e \mathbin{\&\&} e \mid e \mathbin{||} e \mid {!}e & \text{constraints} & \text{\footnotesize\hcode{popcount(?c) == ?s}} \\
  \varnt & ::= & (I_1 \cdots I_m;\; (e_1 @ \lv_1) \cdots (e_r @ \lv_r)) & \text{\variants{}} & \text{\cref{subfig:core-example-surface}} \\
\end{array}
\]
\caption{Abstract Syntax of the Core Language.
$A \rightharpoonup B$ holds the partial maps from $A$ to $B$,
square brackets mark optional parts,
$\langle \eta \times \tau \rangle$ is a vector type,
and $e @ \lv$ is a constraint $e$ with context instruction $\lv$.}
\Description{A grammar table giving the abstract syntax of the MetaphIR core language, with the concept and a concrete syntax example for each syntactic category.}
\label{fig:core-syntax}
\end{figure}

\begin{figure}[t]
\centering
\footnotesize
\renewcommand{\arraystretch}{1}
\newcommand{\rowstrut}{\rule[-1.4ex]{0pt}{4.2ex}}
\begin{subfigure}[t]{0.4\linewidth}
\begin{tabular}{@{}l@{}}
  \rowstrut\footnotesize\hcode{?bb:} \\
  \rowstrut\footnotesize\hcode{\%z = zext:o i8 ?x to i?n} \\
  \rowstrut\footnotesize\hcode{\%a = add:c nsw nuw:?f i?n \%z, ?y} \\
  \rowstrut\footnotesize\hcode{\%c = icmp:?cmp !samesign ?p i?n \%a, 0} \\
  \rowstrut\footnotesize\hcode{;; isKnownNonZero(?y)} \\
\end{tabular}
\caption{Concrete Syntax.}
\label{subfig:core-example-surface}
\end{subfigure}%
\hfill
\begin{subfigure}[t]{0.58\linewidth}
\begin{tabular}{@{}l@{}}
  \rowstrut \\
  \rowstrut$z = \kw{zext}^{\{\mko{\mathit{?x}}\}}_{\emptyset,\,\emptyset}\;\kw{i}\,\mathit{?n}\;
    (\kw{i8}\;\mathit{?x})\;\kwin\;\mathit{?bb}$ \\
  \rowstrut$a = \kw{add}^{\{\mkc\}}_{\{\kw{nsw} \mapsto +,\, \kw{nuw} \mapsto \mathit{?f}\},\,\emptyset}\;
    (\kw{i}\,\mathit{?n}\;z,\; \kw{i}\,\mathit{?n}\;\mathit{?y})\;\kwin\;\mathit{?bb}$ \\
  \rowstrut$c = \kw{icmp}^{\emptyset}_{\{\kw{samesign} \mapsto -\},\,\{\kpred \mapsto \mathit{?p}\}}\;
    (\kw{i}\,\mathit{?n}\;a,\; \kw{i}\,\mathit{?n}\;0)\;\kwas\;\mathit{?cmp}\;\kwin\;\mathit{?bb}$ \\
  \rowstrut$\mathit{isKnownNonZero}(\mathit{?y}) @ c$ \\
\end{tabular}
\caption{Abstract Syntax.}
\label{subfig:core-example-core}
\end{subfigure}
\caption{Concrete and Abstract Syntax of a \Variant{}.}
\Description{A core pattern in concrete syntax on the left and its abstract syntax representation on the right, line by line.}
\label{fig:core-example}
\end{figure}

\subsection{Syntax}
\label{subsec:core-syntax}

We formalize the \emph{core language},
the language of the \variants{} that lowering produces
(\cref{subsec:lowering}).
\Cref{fig:core-syntax} gives its abstract syntax (corresponding to the
IR-like concrete syntax of \cref{subsec:lang-core}),
and \cref{fig:core-example} shows a \variant{} in concrete syntax alongside its
abstract syntax.
A \variant{} $\varnt$ is a sequence of instruction nodes
and a sequence of constraints.
An operand $\lv$ must name an earlier node,
so the nodes form a directed acyclic graph.
The nodes that no other node uses are the \emph{roots} of the \variant{},
and a \variant{} may have more than one root.

\myparagraph{Instruction Nodes}
Each instruction node $I$ defines a local name $\lv$ and has an opcode $\op$,
a flag specification $\flagspec$, keyed arguments $\keyspec$,
an optional type argument $\tau_0$, and typed operands $\tau_i\,o_i$.
A node may also bind a \metavar{} to the instruction that the node matches
($\kwas\;\mvi$)
and require that instruction to lie in a given basic block ($\kwin\;\mvbb$).
Every instruction, whatever its surface syntax, becomes a node of this form.
Most parts of an instruction that are not flags or operands,
such as the predicate of a compare or the alignment of a load,
become keyed arguments.
A type that is not an operand's type,
such as the destination type of a cast,
becomes the type argument.
An operand list may also end with a variadic tail,
which binds the remaining operands and their types to two array \metavars{};
for simplicity, we omit the tail from the rules.

\myparagraph{Flags and Marks}
The flag specification $\flagspec$ maps each flag that the pattern mentions to
$+$ (present), $-$ (absent),
or a \metavar{} that captures whether the flag is present.
A flag that the pattern does not mention may be present or absent.
The marks $\mks$ hold the structural \metaflags{}:
$\mkc$ for commutative matching,
and $\mko{o}$ for a node that may match the operand $o$ instead.

\subsection{Sorts and Well-Formedness}
\label{subsec:core-sorts}

One set of \emph{sorts} classifies both \metavars{} and constraint
expressions:
\[
  \srt ::= \sOpd \mid \sVal \mid \sInst \mid \sFn
    \mid \sInt \mid \sNat \mid \sBool
    \mid \sStr \mid \sPred \mid \sIntr
    \mid \sBB \mid \sType \mid \sBV
    \mid \srt[\,].
\]
They are the sorts of \metavars{} in \cref{tbl:sorts},
the sorts of values that constraints compute
($\sInt$, $\sStr$, and $\sBV$ for bit vectors),
and $\sOpd$, the sort of an operand position.
An operand usually binds an IR value ($\sVal$),
but also a \metavar{} that is a width ($\sNat$), as \code{?n} in
\code{add i?n ?x, ?n},
binds a number and matches the constant with that value.

\myparagraph{Subsorts}
A subsort $\srt \subsort \srt'$ means that every object of sort
$\srt$ is also one of sort $\srt'$,
so such an object may be used wherever $\srt'$ is expected.
The order is generated by
\[
  \sInst, \sFn \subsort \sVal \subsort \sOpd
  \qquad
  \sBool \subsort \sNat \subsort \sOpd
  \qquad
  \sNat \subsort \sInt
  \qquad
  \sPred, \sIntr \subsort \sStr.
\]

\myparagraph{Sorts of \TMetavars{}}
A sort environment $\Gamma$ maps each \metavar{} to the meet
(greatest lower bound) of the sorts that its occurrences require.
In our setting, the meet of two sorts is just the more specific one.
For example, a \metavar{} used as both an operand and a width has sort $\sNat$
(the meet of $\sOpd$ and $\sNat$),
as for \code{?n} in \code{add i?n ?x, ?n}.
When no occurrence is more specific than $\sOpd$,
as for \code{?x} in \code{add i8 ?x, ?y},
$\Gamma$ gives the \metavar{} the default sort $\sVal$.

\myparagraph{Sorts of Constraints}
Each built-in operator has a signature
$f : \srt_1 \times \cdots \times \srt_n \to \srt$,
such as
$\mathit{noOverflow} : \sStr \times \sVal \times \sVal \to \sBool$.
An argument is accepted if its sort lies below a sort that coerces to the
parameter's:
\begin{mathpar}
\inferrule[S-Var]{ }{\Gamma \vdash \mv : \Gamma(\mv)}
\and
\inferrule[S-Loc]{ }{\Gamma \vdash \lv : \sInst}
\and
\inferrule[S-Op]{
  f : \srt_1 \times \cdots \times \srt_n \to \srt \\
  \forall i.\; \Gamma \vdash e_i : \srt'_i \; \wedge \;
  \srt'_i \subsort \srt''_i \coerce \srt_i
}{\Gamma \vdash f(e_1, \ldots, e_n) : \srt}
\end{mathpar}
Binary operators such as \code{+} are just operators in $\Ops$ written infix.
The coercion relation $\coerce$ is reflexive and contains
$\sInt \coerce \sBool$,
$\sVal \coerce \sInt$, $\sType \coerce \sStr$,
$\sVal \coerce \sStr$, $\sBB \coerce \sVal$,
$\sType[\,] \coerce \sStr[\,]$, and $\sVal[\,] \coerce \sStr[\,]$.
A coercion $\cto{c}{\srt}$ converts a value $c$ to sort $\srt$,
or $\err$ if no coercion applies or the conversion fails on $c$.
For example, $\sVal \coerce \sInt$ succeeds only on an integer constant
(\cref{subsec:core-constraints}).

\myparagraph{Well-Formedness}
While parsing, \sys{} checks the surface pattern:
every \metavar{} has a sort,
every constraint has sort $\sBool$,
each flag in $\flagspec$ is supported by $\op$,
the structural mark $\mkc$ appears only on commutative operations
and compares,
and every named \metavar{} is bound by every match.
% even when $\mko{\cdot}$ lets a node match its operand instead.
A pattern that passes these checks is \emph{well-formed}.
% and lowering preserves the checks,
% so every \variant{} is well-formed too.
% The implementation defines the sorts, their order, and the coercions in two
% tables that both the parser and the evaluator use.

\subsection{Targets and Structural Matching}
\label{subsec:core-match}

Structural matching is the judgment $\pmatch{\sigma}{I}{u}$:
the instruction node $I$ of a \variant{} matches the value $u$ of an \llvm{} IR function
under the \instantiation{} $\sigma$.

\begin{table}[t]
\centering
\footnotesize
\caption{Functions on an Instruction $u$.
The examples are for $u$ = \hcode{\%c = icmp samesign ult i32 \%a, \%b} in block
\hcode{\%entry}.}
\label{tbl:inst-parts}
\begin{tabular}{@{}l>{\raggedright\arraybackslash}p{0.5\linewidth}l@{}}
  \toprule
  \textbf{Function} & \textbf{Gives} & \textbf{Example} \\
  \midrule
  $\opcode(u)$ & the opcode & \hcode{icmp} \\
  $\flagsof(u)$ & the set of present flags & $\{\kw{samesign}\}$ \\
  $\keysof(u)$ & the keyed arguments, such as a predicate or an alignment
    & $\{\kpred \mapsto \kw{ult}\}$ \\
  $\tyarg(u)$ & the type argument, usually the result type,
    such as the destination type of a cast & \hcode{i1} \\
  $\blockof(u)$ & the basic block & \hcode{\%entry} \\
  $\opnds(u)$ & the operands $(u_1, \ldots, u_n)$ & $(\hcode{\%a}, \hcode{\%b})$ \\
  $\typeof(u_i)$ & the type of the \textit{i}th operand & \hcode{i32} \\
  \bottomrule
\end{tabular}
\end{table}

\myparagraph{Targets}
A target is an IR function $F$, which we view as a graph
whose nodes are the values of $F$
(its instructions, arguments, constants, globals, and basic blocks)
and whose edges lead from each instruction to its operands.
% Through \code{phi} nodes the graph may contain cycles,
% but patterns are acyclic,
% so matching still terminates.
\Cref{tbl:inst-parts} lists the functions that matching uses to inspect an instruction.

\myparagraph{\Instantiations{}}
An \instantiation{} $\sigma$ maps each local name of $\varnt$ to a value of $F$
and each \metavar{} to a semantic object of its sort:
values for $\sVal$, natural numbers for $\sNat$,
\llvm{} types for $\sType$,
and likewise for the other sorts.

\begin{figure}[t]
\footnotesize
\def\MathparLineskip{\lineskiplimit=0.6em\lineskip=0.4em plus 0.4em}
\mprset{sep=1.4em}
\begin{mathpar}
\inferrule[M-Var]{\sigma(\mv) \doteq u}{\pmatch{\sigma}{\mv}{u}}
\and
\inferrule[M-Const]{k \doteq u}{\pmatch{\sigma}{k}{u}}
\and
\inferrule[M-Loc]{\sigma(\lv) = u}{\pmatch{\sigma}{\lv}{u}}
\\
\inferrule[M-Inst]{
  \opcode(u) = \op \\
  \sigma, \flagsof(u) \fsat \flagspec \\
  \forall \kappa \in \dom(\keyspec).\;
  \sigma(\keyspec(\kappa)) = \keysof(u)(\kappa) \\
  \sigma(\tau_0) \doteq \tyarg(u) \\\\
  \opnds(u) = (u_1, \ldots, u_n) \\
  \forall i.\;
  \sigma(\tau_i) \doteq \typeof(u_i) \wedge \pmatch{\sigma}{o_i}{u_i} \\
  \sigma(\mvi) = u \\
  \sigma(\mvbb) = \blockof(u)
}{\pmatch{\sigma}{\op^{\mks}_{\flagspec,\keyspec}\,\tau_0\,(\tau_1\,o_1, \ldots,
\tau_n\,o_n)\,\kwas\,\mvi\,\kwin\,\mvbb}{u}}
\\
\inferrule[M-Comm]{
  \pmatch{\sigma}{\op^{\mks}_{\flagspec,\keyspec'}\,\tau_0\,(\tau_2\,o_2,
  \tau_1\,o_1)\,\kwas\,\mvi\,\kwin\,\mvbb}{u}
}{\pmatch{\sigma}{\op^{\mks \cup \{\mkc\}}_{\flagspec,\keyspec}\,\tau_0\,
(\tau_1\,o_1, \tau_2\,o_2)\,\kwas\,\mvi\,\kwin\,\mvbb}{u}}
\and
\inferrule[M-Self]{
  \mko{o} \in \mks \\
  \pmatch{\sigma}{o}{u} \\\\
  u \text{ is an instruction} \\
  \sigma(\mvi) = u
}{\pmatch{\sigma}{\op^{\mks}_{\flagspec,\keyspec}\,\tau_0\,(\tau_1\,o_1, \ldots,
\tau_n\,o_n)\,\kwas\,\mvi}{u}}
\\
\inferrule[M-Flags]{
  \forall \fl \in \dom(\flagspec).\;
  \big(\flagspec(\fl) = + \Rightarrow \fl \in S\big) \wedge
  \big(\flagspec(\fl) = - \Rightarrow \fl \notin S\big) \wedge
  \big(\flagspec(\fl) = \mv \Rightarrow (\sigma(\mv) = \true \Leftrightarrow \fl \in
  S)\big)
}{\sigma, S \fsat \flagspec}
\end{mathpar}
\caption{Structural Matching.
The type argument ($\tau_0$), the instruction binding ($\kwas\,\mvi$),
and the block binding ($\kwin\,\mvbb$) are optional;
when they are omitted, the rules drop the premises that check them.
In \textsc{M-Var}, $\doteq$ also lets a $\sNat$ \metavar{} match the constant with its value.
In \textsc{M-Comm}, $\mkc \notin \mks$,
and $\keyspec' = \keyspec$ unless $\op$ is a comparison,
which swaps its predicate with the operands (a predicate \metavar{} is swapped after it is bound).
We write $\sigma(k) = k$ for a constant $k$.
The relation $k \doteq u$ compares integers at the value's width
and also accepts a splat vector of $k$;
type equality $\doteq$ likewise accepts a vector of the scalar type.}
\Description{Inference rules for structural matching: M-Var, M-Const, M-Loc, M-Inst, M-Comm, M-Self, and M-Flags.}
\label{fig:match-rules}
\end{figure}

\myparagraph{Rules}
\Cref{fig:match-rules} gives the rules for $\pmatch{\sigma}{I}{u}$.
\textsc{M-Inst} checks the opcode, the flags, the keyed arguments,
the types, and each operand,
and binds the capture and the block.
The flag premise of \textsc{M-Inst}, $\sigma, \flagsof(u) \fsat \flagspec$,
defined by \textsc{M-Flags},
requires that each $+$ flag is present, each $-$ flag is absent,
and each captured flag binds its \metavar{} to whether it is present.
\textsc{M-Comm} matches a node marked $\mkc$ if the node matches
with its two operands swapped (and $\mkc$ dropped, to avoid swapping again),
and \textsc{M-Self} lets a node marked $\mko{o}$ match a value
whenever $o$ matches it.

\begin{definition}[Structural matches]
\label{def:struct}
Let $F$ be a target and $\varnt$ a \variant{}.
Let $\LNames(\varnt)$ and $\MVars(\varnt)$ be the local names and
the \metavars{} of $\varnt$,
and let $I_\lv$ be the node of $\varnt$ that defines the local name $\lv$.
The set $\Struct(\varnt, F)$ holds the \instantiations{} $\sigma$ with
$\dom(\sigma) = \LNames(\varnt) \cup \MVars(\varnt)$ such
that $\pmatch{\sigma}{I_\lv}{\sigma(\lv)}$ for every $\lv \in
\LNames(\varnt)$.
\end{definition}

\subsection{Constraint Semantics}
\label{subsec:core-constraints}

Constraints evaluate to concrete values $\CVal$.
% mathematical integers, Booleans, strings, \llvm{} types, IR values,
% bit vectors, and arrays of these.
Evaluation can fail, for example on division by zero
or on arithmetic over a non-constant value,
so the result lies in $\CVal_\err = \CVal \uplus \{\err\}$.
% Treating errors as rejection makes patterns robust:
% \code{udiv(?a, ?b) == 1} simply fails to match when \code{?b} is zero,
% as the corresponding C++ guard would.
Each built-in operator $f$ has a concrete semantics
$f_{F, u} : \CVal^n \to \CVal_\err$,
where $F$ is the target function and $u$ is the context instruction,
as some operators rely on information from them.
\Cref{fig:concrete-sem} gives the semantics $\sem{e}_\sigma$.

\begin{figure}[t]
\footnotesize
\renewcommand{\arraystretch}{1.4}
\[
\begin{array}{@{}c@{}}
  \sem{k}_\sigma = k
  \qquad \sem{\mv}_\sigma = \sigma(\mv)
  \qquad \sem{\lv}_\sigma = \sigma(\lv)
  \qquad \sem{{!}e}_\sigma = \neg \cto{\sem{e}_\sigma}{\sBool} \text{ if } \cto{\sem{e}_\sigma}{\sBool} \neq \err, \text{ else } \err \\
  \sem{f(e_1, \ldots, e_n)}_\sigma = f_{F, \sigma(\lv_c)}(\cto{\sem{e_1}_\sigma}{\srt_1}, \ldots, \cto{\sem{e_n}_\sigma}{\srt_n})
  \text{ if no } \cto{\sem{e_i}_\sigma}{\srt_i} = \err, \text{ else } \err \\[4pt]
  \sem{e_1 \mathbin{==} e_2}_\sigma = \begin{cases}
    \sem{e_1}_\sigma = \sem{e_2}_\sigma & \text{if } \sem{e_1}_\sigma, \sem{e_2}_\sigma \in \CVal \text{ have the same sort} \\
    \cto{\sem{e_1}_\sigma}{\srt} = \cto{\sem{e_2}_\sigma}{\srt} & \text{else if } \cto{\sem{e_1}_\sigma}{\srt}, \cto{\sem{e_2}_\sigma}{\srt} \neq \err \text{ for some } \srt \in \{\sInt, \sStr, \sStr[\,]\} \\
    \err & \text{otherwise}
  \end{cases} \\[4pt]
  \sem{e_1 \mathbin{\&\&} e_2}_\sigma = \begin{cases} \false & \text{if } \cto{\sem{e_1}_\sigma}{\sBool} = \false \\ \cto{\sem{e_2}_\sigma}{\sBool} & \text{if } \cto{\sem{e_1}_\sigma}{\sBool} = \true \\ \err & \text{otherwise} \end{cases}
  \qquad
  \sem{e_1 \mathbin{||} e_2}_\sigma = \begin{cases} \true & \text{if } \cto{\sem{e_1}_\sigma}{\sBool} = \true \\ \cto{\sem{e_2}_\sigma}{\sBool} & \text{if } \cto{\sem{e_1}_\sigma}{\sBool} = \false \\ \err & \text{otherwise} \end{cases}
\end{array}
\]
\caption{Concrete Semantics of Constraints for $e @ \lv_c$.
The target function $F$ and the context instruction $\lv_c$ are fixed
and left implicit in $\sem{e}_\sigma$.
Each argument is coerced to the parameter sort of its operator;
$\srt_i$ is the $i$-th parameter sort of $f$,
and $\cto{c}{\srt}$ is the coercion of $c$ to $\srt$
(\cref{subsec:core-sorts}).
Overloaded operators, such as \code{==}, have no fixed parameter sort
and coerce their arguments themselves.
The logical operators short-circuit as in C++.
}
\Description{Equations defining the concrete semantics of constraints: literals, variables, negation, operator application, equality, conjunction, and disjunction.}
\label{fig:concrete-sem}
\end{figure}

\begin{definition}[Matches]
\label{def:match}
The matches of a \variant{} $\varnt$ in a function $F$ are its structural
matches under which every constraint evaluates to $\true$:
\[
  \Match(\varnt, F) = \{\sigma \in \Struct(\varnt, F) \mid \sem{e}_\sigma =
  \true
  \text{ for every constraint } e @ \lv \text{ of } \varnt\}.
\]
% A constraint that evaluates to $\false$ or to $\err$
% thus rejects the structural match.
\end{definition}

\subsection{Lowering}
\label{subsec:lowering}

A surface pattern $\pat$ extends the core language with
definitions of \metaflags{} and \composites{}.
A \metaflag{} $g$ with arguments $\vec a$ on an operand or instruction $x$,
written $x\!:\!g(\vec a)$,
matches $u$ when two conditions hold:
$x$ matches $u$,
and the body of $g$ evaluates to $\true$
once the first parameter of $g$ is substituted by $x$ and the others by $\vec a$.
% with the instruction that $x$ belongs to as the context.
% Here $x$ is a \metavar{}, a local name, or a \composite{} invocation,
% and \code{:or(...)} and \code{:not(...)} combine bodies with $\mathbin{||}$ and ${!}$.
A \composite{} invocation $\lv = m^{\mks}_{\flagspec}\,(\vec a)$
is an instruction node whose opcode is a \composite{} $m$,
with arguments $\vec a$.
It matches
when some branch of $m$ that is compatible with $\flagspec$ matches
after substituting $\vec a$ for the parameters,
and its marks $\mks$ act as on built-in nodes.
% An invocation with \code{:o(N)} also matches whatever its $N$-th operand
% matches,
% as the mark $\mko{\cdot}$ does for a built-in node.
We write $\Match(\pat, F)$ for the resulting set of \instantiations{},
restricted to the \metavars{} that $\pat$ itself declares.
The other \metavars{} of the branch are local to the invocation.
The function $\Lower(\pat)$ produces a finite set of \variants{} by removing both
abstractions, recursively for nested invocations:
\begin{itemize}[leftmargin=*,nosep]
  \item \emph{\Composite{} invocations.}
  An invocation marked \code{:c} is first duplicated with its operands swapped.
  Each branch is inlined as its own \variant{},
  with its \metavars{} renamed apart,
  such as \code{?r\$k} for \code{?k} in the invocation \code{\%r};
  branches whose fixed flags or types conflict with the invocation are dropped.
  % and a flag $\fl$ that the invocation captures as $\mv$ and a branch fixes to $b$
  % becomes the constraint $\mv \mathbin{==} b$.
  A \code{:o(N)} on the invocation marks each branch's root with
  $\mko{o_N}$, where $o_N$ is the invocation's $N$-th operand,
  so the root may match that operand instead.
  \item \emph{\Metaflags{}.}
  Each \metaflag{} becomes its body as a constraint
  whose context is the flagged instruction.
\end{itemize}
Lowering leaves the $\mkc$ and $\mko{\cdot}$ marks of built-in nodes in place,
because \textsc{M-Comm} and \textsc{M-Self} handle them during matching.

\begin{theorem}[Lowering preserves matches]
\label{thm:lowering}
For every well-formed surface pattern $\pat$ and function $F$,
\(
  \Match(\pat, F) \;=\;
  \{\, \sigma|_{\MVars(\pat)} \mid \varnt \in \Lower(\pat),\;
  \sigma \in \Match(\varnt, F) \,\}.
\)
\end{theorem}

\noindent Note that $\sigma|_{\MVars(\pat)}$ restricts $\sigma$ to
the \metavars{} $\MVars(\pat)$ of $\pat$,
because the \variants{} may bind \metavars{} that $\pat$ does not declare,
such as the renamed \metavars{} of inlined branches.
\Cref{app:proof-lowering} proves \cref{thm:lowering} by induction on the depth of \composite{}
definitions.
% The interesting cases are flag compatibility,
% where an invocation's written,
% negated,
% or captured flag must keep exactly the branches whose fixed flags agree with it,
% and hygiene,
% where fresh names guarantee that two invocations of one \composite{} do not
% share
% \metavars{}.
\Cref{thm:lowering} lets the rest of the paper work with \variants{} only.

\section{Pruning Functions by Abstract Interpretation}
\label{sec:abstract-interp}

\sys{} decides whether $\Match(\varnt, F)$ can be non-empty from the
\features{} of $F$ stored in an index
(\StageFilter{} of \cref{fig:overview-example}),
without reading the IR of $F$,
and we prove that this never loses a match.
Throughout this section, we fix an IR function $F$
and leave it implicit in notation such as $\ValsOf(\op, J)$.

\subsection{\Features{} of an IR Function}
\label{subsec:features}

The \features{} $\feat$, an abstraction of $F$, record which values occur where,
which \facts{} hold for them, and how the instructions of $F$ connect
(\cref{fig:features-example} shows an example).

\myparagraph{\Sites{}}
\Features{} describe values by where they occur.
A \emph{\site{}} $s = (\op, J)$ is an opcode $\op$ and a set of slots $J$,
each an operand position or $\res$ for the result.
The values of $F$ at $s$ are
\[
  \ValsOf(\op, J) \;=\;
  \{\, \opnds(u)_j \mid u \in F,\ \opcode(u) = \op,\ j \in J \,\} \;\cup\;
  \{\, u \mid u \in F,\ \opcode(u) = \op,\ \res \in J \,\}.
\]
We write $(\op, j)$ for $(\op, \{j\})$ for simplicity.
In \cref{subfig:features-f},
the value at $(\hcode{urem}, 1)$ is the constant 2,
the value at $(\hcode{urem}, 0)$ is the argument \code{\%x},
and the value at $(\hcode{add}, 0)$ is \code{\%r}.

\myparagraph{\Facts{}}
A \emph{\fact{}} is a unary predicate $p$ on values from a fixed set:
the constant classes such as \code{isPower2}, \code{isa} classes,
\code{hasOneUse},
and equality with some small integers.
% $n \in [-128, 127]$.
% Bounding $n$ keeps the number of \facts{} per \site{} small.
% For a \fact{} that is a built-in operator, such as \code{isPower2},
% $p$ is the operator's concrete semantics
% (\cref{subsec:core-constraints}).
% For the literal $n$, $p(w)$ holds iff $w = n$.
A \site{} records a \fact{} when \emph{some} value there satisfies it:
\begin{equation}
  \label{eq:facts-of}
  \FactsOf(\op, J) \;=\;
  \{\, p \mid \exists w \in \ValsOf(\op, J).\ p(w) = \true \,\}.
\end{equation}
In \cref{subfig:features-f}, $(\hcode{urem}, 1)$ records literal 2, \code{isPower2},
and \code{isKnownNonZero},
but no \code{isZero} (\cref{subfig:features-table}).
% A recorded \fact{} thus says only that some value at the \site{}
% satisfies it,
% but a \emph{missing} \fact{} says that \emph{no} value there does.
Pruning relies on the missing \facts{}:
if a pattern requires a value with $p$ at a \site{}
and $p$ is missing there,
the function has no match,
as for the first query in \cref{subfig:features-queries}.
Two further kinds of \facts{} are recorded in the same way.
First, a value-tracking predicate $p$ such as \code{isKnownNonZero}
depends on a context instruction,
so $p \in \FactsOf(\op, j)$ if $p$ holds for $\opnds(u)_j$ at the context $u$
itself, for some $u$ with opcode $\op$.
In \cref{subfig:features-f}, value tracking knows that 2 is non-zero at \code{\%r},
so $(\hcode{urem}, 1)$ records \code{isKnownNonZero}.
Second, queries that compare a numeric property with a threshold,
such as the number of known trailing zero bits,
use a \emph{\measure{}}, a \fact{} that carries a number:
a \site{} records the minimum or maximum of the property over its values,
and a query whose threshold lies beyond it is ruled out.

\myparagraph{Structural \Features{}}
Finally,
$\feat$ also records the opcodes present in $F$,
the \emph{edges} $(\op_1, \op_2)$ such that an instruction with opcode $\op_1$
uses one
with opcode $\op_2$, the \emph{siblings} $(\op_1, \op_2)$
whose instructions share an operand, the globals used,
and a partial \emph{signature} for each instruction.
In \cref{subfig:features-f}, these are the opcodes \code{urem} and \code{add},
the edge $(\hcode{add}, \hcode{urem})$ because \code{\%s} uses \code{\%r},
and the signatures \code{urem i32} and \code{add i32}.
% its opcode, flags,
% and result and operand types,
% of which only the first few operand types are kept.

\begin{figure}[t]
  \footnotesize
  \begin{subfigure}[b]{0.34\linewidth}
\begin{lstlisting}[style=llvmstyle,escapechar={},xleftmargin=0pt,basicstyle=\footnotesize\ttfamily]
define i32 @f(i32 %x, i32 %y) {
  %r = urem i32 %x, 2
  %s = add i32 %r, %y
  ret i32 %s
}
\end{lstlisting}
  \caption{The Function \code{f}.}
  \label{subfig:features-f}
  \end{subfigure}%
  \hfill
  \begin{subfigure}[b]{0.64\linewidth}
  \centering
  \begin{tabular}{@{}l>{\raggedright\arraybackslash}p{0.7\linewidth}@{}}
    \toprule
    \textbf{\Site{}} & \textbf{Recorded \Features{}} \\
    \midrule
    $(\hcode{urem}, 1)$ & literal 2, \hcode{isPower2}, \hcode{isStrictlyPositive}; \hcode{isKnownNonZero} \\
    $(\hcode{urem}, 0)$ & none (\hcode{\%x} is an argument) \\
    $(\hcode{add}, 0)$ & range fits in one bit (\hcode{\%r} is 0 or 1) \\
    structural & opcodes \hcode{urem}, \hcode{add}; edge $(\hcode{add}, \hcode{urem})$; \newline
    signatures \hcode{urem i32}, \hcode{add i32} \\
    \bottomrule
  \end{tabular}
  \caption{Some \Features{} of \code{f}.}
  \label{subfig:features-table}
  \end{subfigure}
  \par\medskip
  \begin{subfigure}{\linewidth}
  \centering
  \begin{tabular}{@{}lll@{}}
    \toprule
    \textbf{Query} & \textbf{Structural Requirements} & \textbf{Why \code{f} Is Ruled Out} \\
    \midrule
    \begin{tabular}[t]{@{}l@{}}\hcode{\%r = urem i32 ?x, ?c} \\ \hcode{;; isZero(?c)}\end{tabular} & opcode \hcode{urem}, width 32 & \hcode{isZero} is not recorded at $(\hcode{urem}, 1)$ \\
    \midrule
    \begin{tabular}[t]{@{}l@{}}\hcode{\%t = add i32 ?a, ?b} \\ \hcode{\%u = urem i32 \%t, ?c}\end{tabular} & \begin{tabular}[t]{@{}l@{}}opcodes \hcode{add}, \hcode{urem}, width 32,\\ edge $(\hcode{urem}, \hcode{add})$\end{tabular} & no edge $(\hcode{urem}, \hcode{add})$ \\
    \bottomrule
  \end{tabular}
  \caption{Two Queries That the \Features{} Rule Out,
  with Their Structural Requirements (\cref{subsec:requirements}).}
  \label{subfig:features-queries}
  \end{subfigure}
  \caption{\Features{} of a Function and Their Use for Pruning.}
  \Description{A small LLVM function, a table of some of its recorded features, and two example queries that the features rule out, with the reason for each.}
  \label{fig:features-example}
\end{figure}

\subsection{Structural Requirements of a \Variant{}}
\label{subsec:requirements}

\Features{} summarize an IR function,
while a \variant{} $\varnt$'s structural requirements specify the
structural \features{} a function must have to match it.
The \emph{certain nodes} of $\varnt$ are the nodes
without an $\mko{\cdot}$ mark,
which every match maps to an instruction with their own opcode.
The \emph{structural requirements} $\Req(\varnt)$ of $\varnt$ are
\circled{1}~the opcode of each certain node,
\circled{2}~the edge $(\op_1, \op_2)$ whenever a certain node with opcode $\op_1$
has a certain operand node with opcode $\op_2$,
\circled{3}~the sibling pair $(\op_1, \op_2)$ whenever a \metavar{} is an operand
of two certain nodes with these opcodes,
\circled{4}~the written and negated flags and the fixed widths of each certain node,
and \circled{5}~each global that $\varnt$ names.
The second query in \cref{subfig:features-queries} requires the edge
$(\hcode{urem}, \hcode{add})$,
which \cref{subfig:features-table} lacks,
so \code{f} is ruled out for this query by \cref{lem:requirements} below.

\begin{lemma}[Requirements are necessary]
\label{lem:requirements}
If $\Match(\varnt, F) \neq \emptyset$, then $\Req(\varnt) \subseteq \feat$.
\end{lemma}

\myparagraph{Typings}
Type-level \metavars{} such as \code{?bits} can be bound from the signatures.
A \emph{typing} is a partial map from type-level \metavars{} to \llvm{}
types and sizes.
The set $\JoinSig(\varnt)$ holds the typings of $\varnt$ that the
partial signatures in $\feat$ allow.
For example, for a \variant{} with nodes \hcode{urem i?n ...} and
\hcode{add i?n ...},
the signatures \code{urem i32} and \code{add i32} in
\cref{subfig:features-table}
give it the single typing $\code{?n} = 32$.
Because each node may match multiple signatures in $\feat$,
the number of combinations can grow exponentially with the number of nodes;
\sys{} therefore caps $\JoinSig(\varnt)$ at \MaxTypings{} typings.
Above the cap, \sys{} uses the empty typing.%, which is sound but prunes less.

\begin{lemma}[Typings are sound]
\label{lem:typings}
If $\sigma \in \Match(\varnt, F)$,
then some $\theta \in \JoinSig(\varnt)$ satisfies
$\theta \subseteq \sigma$.
\end{lemma}

\subsection{Evaluating Constraints over an Abstract Domain}
\label{subsec:evaluator}

Constraints share a single evaluator parameterized by a value domain:
\StageEval{} runs it on concrete IR values ($\sem{\cdot}_\sigma$, \cref{fig:concrete-sem}),
and \StageFilter{} runs it on the \emph{index domain} over \features{} (\cref{fig:evaluator}).
Both stages thus share expression traversal, coercions, and concrete operator evaluation,
reducing the soundness of the abstract constraint evaluation to the soundness of each \abstractFace{}.

\myparagraph{Abstract Values}
Fix a \variant{} $\varnt$, a function $F$, and a typing $\theta$.
An abstract value describes the results that an expression can have across all
matches of $\varnt$ in $F$ that agree with $\theta$.
\Cref{subfig:abstract-values} lists the four kinds of abstract values,
which form the abstract domain $\AVal$,
and their concretization $\gam : \AVal \to \CDom$
into the concrete domain $\CDom = \mathcal{P}(\CVal_\err)$.
% Every concretization contains $\err$,
% so the domain need not track errors,
% which can arise in any match, for example from a failed coercion.
% Including $\err$ costs no pruning,
% because an error rejects a match just as $\false$ does:
% $\conc{\false}$ means ``false or an error'',
% and either way the match is rejected.

\myparagraph{\Sites{} of a \TMetavar{} or Local Name}
The set $\SitesOf(x)$ holds the \sites{} where a \metavar{} or local name $x$
certainly occurs in every match.
If $\mv$ is operand $j$ of a certain node with opcode $\op$,
then $(\op, \{j\}) \in \SitesOf(\mv)$,
or $(\op, \{0, 1\})$ when the node has the $\mkc$ mark,
since $\mv$ may be in either slot.
A local name $\lv$ and its capture get the \site{} $(\op, \{\res\})$
if their node is certain,
and also the \site{} of $\lv$ as an operand of each certain user.
For example, in
\code{\%r = add i32 ?a, 1} and \code{\%s = urem i32 \%r, ?c},
$\SitesOf(\code{\%r}) = \{(\code{add}, \{\res\}), (\code{urem}, \{0\})\}$.
If \code{add} has mark \code{:o}, $\SitesOf(\code{\%r})$ loses $(\code{add}, \{\res\})$.
% because it may match \code{?a} instead of an \code{add}.

\begin{figure}[t]
\footnotesize
\begin{subfigure}[b]{0.6\linewidth}
\centering
\begin{tabular}{@{}lll@{}}
  \toprule
  \textbf{Abstract value} & \textbf{Meaning} & \textbf{Concretization $\gam$} \\
  \midrule
  $\conc{c}$ & the constant $c$ & $\{c, \err\}$ \\
  $\sitesv{S}$ & at every \site{} in $S$ & $\bigcap_{s \in S} \ValsOf(s) \cup \{\err\}$ \\
  $\err$ & an error & $\{\err\}$ \\
  $\unk$ & unknown & $\CVal_\err$ \\
  \bottomrule
\end{tabular}
\caption{Abstract Values $\AVal$.}
\label{subfig:abstract-values}
\end{subfigure}%
\hfill
\begin{subfigure}[b]{0.38\linewidth}
\centering
$\begin{array}{@{}l@{}}
  \mathbb{F} = \{\conc{\false}, \err\} \\
  \mathbb{B} = \mathbb{F} \cup \{\conc{\true}\} \\
  \fn{lift}(c) = \conc{c},\ \fn{lift}(\err) = \err \\
  \asem{\vec e} = \asem{e_1}, \ldots, \asem{e_n} \\
  \hat{f} = \unk \text{ if } f \text{ has no \abstractFace{}}
\end{array}$
\caption{Notation.}
\label{subfig:abstract-notation}
\end{subfigure}

\vspace{1ex}
\begin{subfigure}{\linewidth}
\[
\begin{array}[t]{@{}r@{\;}c@{\;}l@{\;}l@{}}
  \asem{k} & = & \conc{k} \\[0.5ex]
  \asem{\mv} & = & \conc{\theta(\mv)} & \text{if } \mv \in \dom(\theta) \\
  & & \sitesv{\SitesOf(\mv)} & \text{if } \SitesOf(\mv) \neq \emptyset \\
  & & \unk & \text{otherwise} \\[0.5ex]
  \asem{f(\vec e)} & = & \fn{lift}(f(\vec c)) & \text{if each } \asem{e_i} = \conc{c_i} \\
  & & \hat{f}(\asem{\vec e}) & \text{otherwise} \\[0.5ex]
  \asem{{!}e} & = & \conc{\neg b} & \text{if } \asem{e} = \conc{b} \\
  & & \err & \text{if } \asem{e} = \err \\
  & & \unk & \text{otherwise}
\end{array}
\;\;
\begin{array}[t]{@{}r@{\;}c@{\;}l@{\;}l@{}}
  \asem{e_1 \mathbin{\&\&} e_2} & = & \asem{e_1} & \text{if } \asem{e_1} \in \mathbb{F} \\
  & & \asem{e_2} & \text{if } \asem{e_1} = \conc{\true} \\
  & & \conc{\false} & \text{if } \asem{e_1} \notin \mathbb{B},\, \asem{e_2} \in \mathbb{F} \\
  & & \unk & \text{otherwise} \\[0.5ex]
  \asem{e_1 \mathbin{||} e_2} & = & \asem{e_1} & \text{if } \asem{e_1} \in \{\conc{\true}, \err\} \\
  & & \asem{e_2} & \text{if } \asem{e_1} = \conc{\false} \\
  & & \conc{\true} & \text{if } \asem{e_1} \notin \mathbb{B},\, \asem{e_2} = \conc{\true} \\
  & & \unk & \text{otherwise}
\end{array}
\]
\caption{The Abstract Semantics $\asem{e}_\theta$ of Constraints.}
\label{subfig:abstract-sem}
\end{subfigure}
\caption{The Index Domain.
In \subref{subfig:abstract-sem}, we omit the subscript $\theta$,
local names evaluate as \metavars{} do,
and constants are coerced to $\sBool$ as in \cref{fig:concrete-sem}.}
\Description{A table of the four abstract values and their concretizations, a notation box, and equations for the abstract semantics of constraints.}
\label{fig:evaluator}
\end{figure}

\myparagraph{Abstract Semantics}
The \emph{abstract semantics} $\asem{e}_\theta$ evaluates a constraint $e$
in the index domain under the typing $\theta$
(\cref{subfig:abstract-sem}).
It differs from \cref{fig:concrete-sem} in only three places.
A variable evaluates to what the \features{} know about it.
An operator evaluates concretely when all its operands are concrete,
and otherwise asks the operator's \emph{\abstractFace{}} $\hat{f}$.
Finally, a connective whose left side is unknown may still be decided by its
right side, as in Kleene logic.
% Because both instances share $\Eval$,
% the two instances cannot disagree about short-circuiting, coercions,
% or the meaning of an operator on concrete operands:
% these come from the same code.

\myparagraph{\TAbstractFaces{}}
An operator's \abstractFace{} returns $\conc{\false}$ only
when a recorded \feature{} rules out every value,
and $\unk$ otherwise.
For a \fact{} predicate $p$ (including $\mv = n$ for a small literal $n$),
its \abstractFace{} $\hat{p}$ gives $\hat{p}(\sitesv{S}) = \conc{\false}$ if $p \notin \FactsOf(s)$ for some
$s \in S$;
a threshold query likewise returns $\conc{\false}$ when its threshold lies beyond the \measure{} at $s$.
For a value-tracking predicate $p$ in $e @ \lv_c$,
the index computes $p$ only with the instruction that uses the operand as the context (\cref{subsec:features});
so $\hat{p}$ checks only the \site{} where the argument is an operand of $\lv_c$ itself,
when $\lv_c$ is certain,
and returns $\unk$ otherwise.
% \begin{itemize}[leftmargin=*,nosep]
%   \item a \fact{} predicate $p$,
%   including $\mv = n$ for a small literal $n$:
%   $\hat{p}(\sitesv{S}) = \conc{\false}$ if $p \notin \FactsOf(s)$ for some
%   $s \in S$;
%   \item a value-tracking predicate $p$ at context $\lv_c$:
%   $\conc{\false}$ if $\lv_c$ is certain and $p \notin \FactsOf(s)$ for some
%   \site{} $s \in S$ that $\mv$ occupies as an operand of $\lv_c$ itself;
%   \item a threshold query with \measure{} $\nu$ and a concrete threshold $t$:
%   $\conc{\false}$ if $t$ lies beyond the \measure{} of $\nu$
% at such a \site{} $s$.
% \end{itemize}
Each \abstractFace{} is sound in the following sense:
if $\hat{f}(\vec{\hat a}) = \conc{\false}$ and $w_i \in \gam(\hat a_i)$,
then $f_{F,u}(\vec w) \in \{\false,
\err\}$ for the context $u$ of any match.
For a \fact{} predicate $p$,
this holds because $w \in \ValsOf(s)$ and $p \notin \FactsOf(s)$ give
$p(w) \neq \true$ by~\cref{eq:facts-of}.
\Cref{tbl:ai-examples} evaluates sample constraints on function \code{f}.

\begin{table}[t]
  \footnotesize
  \centering
  \caption{Abstract Evaluation on Function \code{f} of
  \cref{subfig:features-f}, for the \Variant{} \hcode{\%r = urem i?n ?x, ?c}
  and the Typing $\code{?n} = 32$.}
  \label{tbl:ai-examples}
  \begin{tabular}{@{}lll@{}}
    \toprule
    \textbf{Constraint} & \textbf{Abstract result} & \textbf{Function} \\
    \midrule
    \hcode{?c == 2} & $\unk$: literal 2 is recorded at $(\hcode{urem}, 1)$, so \hcode{?c} may be 2 & kept \\
    \hcode{?c == 4} & $\conc{\false}$: literal 4 is missing & pruned \\
    \hcode{?c == 4 || ?c == 2} & $\conc{\false} \vee \unk = \unk$ & kept \\
    \hcode{?c == ?x \&\& ?c == 4} & $\unk \wedge \conc{\false} = \conc{\false}$ & pruned \\
    \hcode{isKnownNonZero(?x)} & $\conc{\false}$: value-tracking \fact{} missing at $(\hcode{urem}, 0)$ & pruned \\
    \hcode{isZero(add(?c, 1))} & $\unk$: \hcode{add} has no \abstractFace{} & kept \\
    \hcode{?n == 8} & $\conc{32 == 8} = \conc{\false}$ & pruned \\
    \hcode{udiv(1, 0) == 0} & $\err$ & pruned \\
    \bottomrule
  \end{tabular}
\end{table}

\subsection{Soundness}
\label{subsec:soundness}

The following lemma states that the index domain over-approximates the concrete
domain at every subexpression.

\begin{lemma}[Abstract evaluation is sound]
\label{lem:eval-sound}
Let $\sigma \in \Struct(\varnt, F)$ and $\theta \subseteq \sigma$.
For every constraint $e @ \lv_c$ of $\varnt$
and every subexpression $e'$ of $e$,
$\sem{e'}_\sigma \in \gam(\asem{e'}_\theta)$.
\end{lemma}

The proof is by induction on $e'$ (see \cref{app:proof-soundness}).
% variables follow from $\theta \subseteq \sigma$
% and from $\sigma(\mv) \in \ValsOf(s)$ for each \site{} $s$ of $\mv$,
% operators from the soundness of the \abstractFaces{},
% and the connectives by case analysis on \cref{fig:evaluator},
% using that $\err$ lies in every concretization.

\begin{theorem}[Pruning is sound]
\label{thm:soundness}
Let $\varnt$ be a \variant{}.
If $\Req(\varnt) \not\subseteq \feat$,
or if for every typing $\theta \in \JoinSig(\varnt)$ some constraint
$e$ of $\varnt$ has $\asem{e}_\theta \in \{\conc{\false},
\err\}$, then $\Match(\varnt, F) = \emptyset$.
\end{theorem}

\begin{proof}
Suppose $\sigma \in \Match(\varnt, F)$.
By \cref{lem:requirements}, $\Req(\varnt) \subseteq \feat$.
By \cref{lem:typings},
some $\theta \in \JoinSig(\varnt)$ has $\theta \subseteq \sigma$;
take the constraint $e$ that the hypothesis gives for it.
By \cref{lem:eval-sound},
$\sem{e}_\sigma \in \{\false, \err\}$,
so $\sigma \notin \Match(\varnt, F)$, a contradiction.
\end{proof}

\begin{corollary}[Indexed search is exact]
\label{cor:exact}
For every pattern and corpus,
pruning with the index before matching the functions
returns the same matches as matching every function in the corpus.
\end{corollary}

The corollary follows from \cref{thm:lowering,thm:soundness}.

\section{Implementation}
\label{sec:impl}

\sys{} is implemented in C++ on top of \llvm{}.
It consists of a parser, the lowering pass, the matcher,
the constraint evaluator with its two domains,
and the corpus index, exposed through command-line tools
and a WebAssembly-based application.

\myparagraph{The Corpus Index}
In addition to the posting lists,
the index stores \features{} at two levels:
per-module \features{} (the union of its functions' \features{})
and the full \features{} of each function, decoded only on demand.
A query intersects the shortest few posting lists of each \variant{},
takes the union over \variants{}, checks the per-module \features{},
and only then decodes the \features{} of each candidate
and runs the abstract evaluation.
% Value-tracking \facts{} are the most expensive to compute,
% so they are recorded for at most \ValueFactBudget{} values per function, beyond
% which the function's value-tracking \facts{} are treated as unknown.
Over the evaluation corpus, the index takes \IndexSizeGB{}~GB,
against \CorpusSizeGB{}~GB of bitcode.

\myparagraph{Web Application}
The C++ query engine also compiles to WebAssembly (\WasmSizeMB{}~MB).
A web application lets a user write a pattern
and run it against the published index (\WebIndexSizeMB{}~MB).
Alongside the index, each IR function is packed into a bitcode module (\PacksSizeGB{}~GB).
The application reads the index and the packed function IR with HTTP range
requests,
so a query downloads only the posting lists and candidate functions it needs.

% \myparagraph{Testing}
% \sys{} has \NumUnitTests{} unit tests and \NumLitTests{} end-to-end regression
% tests.
% Every unit test of the feature filter checks \cref{cor:exact} directly:
% it matches with and without pruning, with and without value-tracking \facts{},
% and requires equal results.
% A differential test runs every regression test in \NumDiffModes{} modes (no
% pruning, prefiltering, parallel, indexed, trusting the index, and packed IR)
% and requires identical output.

\section{Evaluation}
\label{sec:eval}

We evaluate \sys{} on \NumPRsWord{} real \instcombine{} rewrites
and a large corpus of IR compiled from real-world programs,
and answer three research questions.
\begin{itemize}[leftmargin=*,nosep]
  \item \textbf{RQ1 (Expressiveness).}
  Can the \sys{} \dsl{} express where \llvm{} rewrites apply?
  \item \textbf{RQ2 (Accuracy).}
  How accurately do the matches identify where the rewrites apply,
  and how does this compare with \daggrep{}?
  \item \textbf{RQ3 (Efficiency).}
  How much work does the index save, and how fast are queries?
\end{itemize}

\subsection{Setup}
\label{subsec:eval-setup}

We describe the rewrites, the datasets, the \gt{}, and the machine.

\myparagraph{Rewrites}
We collected \NumPRsWord{} \llvm{} PRs (\cref{tbl:accuracy}) that each add
or generalize an \instcombine{} or \instsimplify{} rewrite.
They were merged between \PRMergedFrom{} and \PRMergedTo{}.
% and were written by \NumPRAuthors{} different contributors.
Each PR states a clear transformation from a source \src{} to a target \tgt{}
and has a regression test.
The rewrites cover arithmetic, bitwise, shift, compare, select, and cast operations.
For each PR,
we wrote one or two patterns specifying where the rewrite applies,
totaling \NumPatterns{} patterns.

\myparagraph{Datasets}
Our main dataset is \corpusName{}~\cite{llvm_opt_benchmark},
the corpus \llvm{} developers use to assess the applicability of optimizations.
It holds the \llvmir{} of \NumProjects{} open-source projects in C, C++,
and Rust, including the Linux kernel, FFmpeg, and OpenCV,
in \NumBenchModules{} modules.
To obtain IR on which the rewrites can still apply,
we built an \llvm{} with each of the \NumPRsWord{} rewrites disabled
and ran its \code{O2} pipeline on every module.
We drop the functions marked \code{optnone} (disabled optimizations),
since the compiler does not optimize them.
As a second dataset,
we used the upstream \instcombine{} and \instsimplify{} test suites
(\NumTestFiles{} files).
We evaluate accuracy separately on the two datasets (RQ2).
The index and timing measurements (RQ3) are one run over both,
\NumFunctions{} functions and \CorpusSizeGB{}~GB of bitcode,
of which the test suites are a negligible share.

\myparagraph{Ground Truth}
We extract each rewrite's C++ matching logic verbatim into a standalone \llvm{}
pass plugin, replacing the IR transformation with a logging call.
The functions where this plugin logs a match form the \gt{} $\mathit{GT}$,
and the functions matched by a \sys{} pattern form the predictions $P$.
We compare them as sets of (module, function) pairs
and count the true positives $\mathit{TP} = |\mathit{GT} \cap P|$.
We report precision $\mathit{TP} / |P|$, recall $\mathit{TP} / |\mathit{GT}|$,
and their harmonic mean F1.

\myparagraph{Machine}
All experiments ran on a machine with an 80-core Neoverse-N1 CPU (aarch64),
502~GiB of memory, and Ubuntu 22.04,
using \NumThreads{} threads.
Timings are the median of three runs.

\subsection{RQ1: Expressiveness}
\label{subsec:rq1}

The \sys{} \dsl{} expresses where the rewrite applies in each of the \NumPRsWord{} PRs,
aside from a single corner case in the test suites (\cref{subsec:rq2}).
The \NumPatterns{} patterns are compact, totaling \NumPatternInsts{} instruction
lines and \NumPatternConstraints{} constraint lines, with at most
\MaxPatternInsts{} instructions per pattern.
In contrast, \daggrep{} checks only instruction opcodes, flags, literal
constants, \cpp{m_ImmConstant}, and \cpp{isGuaranteedNotToBePoison}, and must
drop single-use checks, comparison predicates, constant classes, and
other constraints.
As a result, it captures all preconditions of only \DaggrepExpressible{} of the
\NumPatterns{} patterns, namely \llvmPR{204089}, \llvmPR{207564}, and
\llvmPR{213313}, and even on \llvmPR{207564} it overmatches because it ignores
types.
It also requires \DaggrepInsts{} instruction lines against \NumPatternInsts{} in
\sys{}, because each alternative of a \composite{} must be written as a separate
pattern file (see \cref{app:expressiveness}).

Beyond the evaluated rewrites,
the standard library of \sys{} reproduces \NumStdComposites{} \patternmatch{} matchers
that combine several instructions (such as \cpp{m_AddLike}, \cpp{m_Signum}, and
\cpp{m_LogicalAnd}), all \NumConstClasses{} constant classes,
and \NumValueTrackOps{} \valuetracking{} queries.
To measure coverage beyond these rewrites,
we counted the \patternmatch{} matchers and \valuetracking{} functions
that \instcombine{} calls at \llvm{} commit \LlvmCommit{}, the one we evaluated on.
A matcher counts as covered if a pattern can express it
with an instruction, a flag, a constant class, a standard-library
\composite{}, or a literal.
\instcombine{} uses \NumPMUsed{} distinct matchers,
and \sys{} covers \NumPMCovered{} of them (\PMCoveredPct{}),
which account for \PMCallsCoveredPct{} of the \NumPMCalls{} calls.
% The uncovered ones are mostly floating-point constants, shuffle masks,
% and \cpp{m_ImmConstant}.
A \cpp{match} call combines several matchers, so it is expressible only if all of them are covered.
By this measure, \NumMatchExprsCovered{} of the \NumMatchExprs{} \cpp{match} calls (\MatchExprsCoveredPct{}) are expressible.
% and \cpp{m_ImmConstant} alone rules out \NumImmConstantBlocked{} of the others.
Of the \NumVTUsed{} \valuetracking{} functions that \instcombine{} calls,
\sys{} covers \NumVTCovered{} (\VTCoveredPct{}),
which account for \VTCallsCoveredPct{} of the \NumVTCalls{} calls.
Supporting the remaining matchers and \valuetracking{} functions is primarily a matter of engineering effort rather than a design limitation, and is left to future work.

\finding{The \sys{} \dsl{} expresses all \NumPRsWord{} rewrites
more compactly and completely than \daggrep{} (\NumPatternInsts{} vs.\ \DaggrepInsts{} instruction lines,
whereas \daggrep{} expresses only \DaggrepExpressible{} of them),
and covers \MatchExprsCoveredPct{} of \instcombine{}'s match expressions.}

\begin{table}[t]
  \centering
  \caption{Per-Rewrite Accuracy and Query Time (seconds) on \corpusName{}.
  $|\mathit{GT}|$, $|P|$, and TP count the functions in the ground truth
  (where the C++ rewrite applies), in the prediction (matched by a tool),
  and in both (true positives, $\mathit{GT} \cap P$).
  P/R/F1 are in percent; a precision or recall with an empty denominator counts as 100.
  The last two rows give the micro and macro averages of P/R/F1,
  and the summed counts and median times and sizes.
  $T$ is the query time, and $T_{\text{NC}}$ and $T_{\text{NI}}$ are the \sys{} times without abstract constraint evaluation (over the
  \features{}) and without the index.
  Web gives the downloaded size (Sz, in MB) and the time in the web application at 10 MB/s, with the full index and without abstract constraint evaluation (NC).}
  \label{tbl:accuracy}
  \footnotesize
  \setlength{\tabcolsep}{3pt}
  \resizebox{\linewidth}{!}{%
  \begin{tabular}{@{}lrrrr@{\,/\,}r@{\,/\,}rrrrrrr@{\,/\,}r@{\,/\,}rrrrrr@{}}
    \toprule
    & & \multicolumn{8}{c}{\textbf{\sys{}}} & \multicolumn{6}{c}{\textbf{\daggrep{}}} & \multicolumn{4}{c}{\textbf{Web}} \\
    \cmidrule(lr){3-10}\cmidrule(lr){11-16}\cmidrule(lr){17-20}
    \textbf{PR} & \textbf{$|\mathit{GT}|$} & \textbf{$|P|$} & \textbf{TP} & \multicolumn{3}{c}{\textbf{P/R/F1}} & \textbf{$T$} & \textbf{$T_{\text{NC}}$} & \textbf{$T_{\text{NI}}$} & \textbf{$|P|$} & \textbf{TP} & \multicolumn{3}{c}{\textbf{P/R/F1}} & \textbf{$T$} & \textbf{Sz} & \textbf{$T$} & \textbf{Sz$_{\text{NC}}$} & \textbf{$T_{\text{NC}}$} \\
    \midrule
    \llvmPRNum{204089} & 0 & 0 & 0 & 100.0 & 100.0 & 100.0 & 0.9 & 0.9 & 26.2 & 0 & 0 & 100.0 & 100.0 & 100.0 & 43.1 & 97 & 10.8 & 131 & 14.4 \\
    \llvmPRNum{204927} & 1,973 & 1,973 & 1,973 & 100.0 & 100.0 & 100.0 & 3.3 & 3.2 & 26.0 & 4,343 & 1,973 & 45.4 & 100.0 & 62.5 & 64.5 & 1,825 & 283.5 & 2,194 & 358.5 \\
    \llvmPRNum{204928} & 1,266 & 1,266 & 1,266 & 100.0 & 100.0 & 100.0 & 1.5 & 1.7 & 26.0 & 5,163 & 1,266 & 24.5 & 100.0 & 39.4 & 65.1 & 343 & 56.2 & 370 & 65.5 \\
    \llvmPRNum{205383} & 0 & 0 & 0 & 100.0 & 100.0 & 100.0 & 0.9 & 1.0 & 51.5 & 0 & 0 & 100.0 & 100.0 & 100.0 & 86.6 & 70 & 8.4 & 72 & 8.6 \\
    \llvmPRNum{207462} & 31 & 31 & 31 & 100.0 & 100.0 & 100.0 & 2.3 & 2.4 & 51.3 & 212 & 31 & 14.6 & 100.0 & 25.5 & 42.5 & 826 & 87.8 & 885 & 93.9 \\
    \llvmPRNum{207564} & 52 & 52 & 52 & 100.0 & 100.0 & 100.0 & 1.4 & 1.6 & 26.3 & 54 & 52 & 96.3 & 100.0 & 98.1 & 21.7 & 539 & 57.3 & 736 & 78.2 \\
    \llvmPRNum{209322} & 0 & 0 & 0 & 100.0 & 100.0 & 100.0 & 0.9 & 4.4 & 25.9 & 68,175 & 0 & 0.0 & 100.0 & 0.0 & 21.4 & 852 & 89.7 & 3,564 & 382.6 \\
    \llvmPRNum{210499} & 10 & 10 & 10 & 100.0 & 100.0 & 100.0 & 2.0 & 2.0 & 26.8 & 5,524 & 10 & 0.2 & 100.0 & 0.4 & 86.7 & 1,169 & 125.9 & 1,925 & 206.6 \\
    \llvmPRNum{212132} & 0 & 0 & 0 & 100.0 & 100.0 & 100.0 & 3.2 & 3.4 & 51.7 & 0 & 0 & 100.0 & 100.0 & 100.0 & 44.1 & 812 & 85.5 & 852 & 89.6 \\
    \llvmPRNum{213313} & 0 & 0 & 0 & 100.0 & 100.0 & 100.0 & 0.8 & 0.8 & 26.1 & 0 & 0 & 100.0 & 100.0 & 100.0 & 43.1 & 68 & 7.6 & 68 & 7.6 \\
    \llvmPRNum{213709} & 9 & 9 & 9 & 100.0 & 100.0 & 100.0 & 1.8 & 2.5 & 26.0 & 9 & 9 & 100.0 & 100.0 & 100.0 & 43.5 & 863 & 91.3 & 1,256 & 132.8 \\
    \llvmPRNum{214376} & 0 & 0 & 0 & 100.0 & 100.0 & 100.0 & 1.2 & 1.2 & 25.9 & 0 & 0 & 100.0 & 100.0 & 100.0 & 21.1 & 151 & 16.2 & 226 & 24.0 \\
    \llvmPRNum{214381} & 9 & 9 & 9 & 100.0 & 100.0 & 100.0 & 0.7 & 1.0 & 25.3 & 14 & 9 & 64.3 & 100.0 & 78.3 & 21.5 & 183 & 19.4 & 223 & 23.5 \\
    \llvmPRNum{214517} & 24 & 24 & 24 & 100.0 & 100.0 & 100.0 & 0.9 & 1.2 & 26.7 & 475 & 24 & 5.1 & 100.0 & 9.6 & 21.4 & 269 & 28.4 & 289 & 30.6 \\
    \llvmPRNum{214541} & 18 & 18 & 18 & 100.0 & 100.0 & 100.0 & 0.6 & 0.7 & 26.3 & 284 & 18 & 6.3 & 100.0 & 11.9 & 20.7 & 73 & 8.2 & 81 & 9.0 \\
    \llvmPRNum{215122} & 32 & 32 & 32 & 100.0 & 100.0 & 100.0 & 2.0 & 2.9 & 26.0 & 168 & 32 & 19.0 & 100.0 & 32.0 & 21.9 & 907 & 95.5 & 1,483 & 156.3 \\
    \llvmPRNum{215130} & 3 & 3 & 3 & 100.0 & 100.0 & 100.0 & 0.9 & 0.9 & 26.6 & 14 & 3 & 21.4 & 100.0 & 35.3 & 22.4 & 52 & 5.9 & 55 & 6.2 \\
    \llvmPRNum{216046} & 57 & 57 & 57 & 100.0 & 100.0 & 100.0 & 1.4 & 1.7 & 26.5 & 1,680 & 57 & 3.4 & 100.0 & 6.6 & 21.2 & 233 & 25.0 & 274 & 29.3 \\
    \llvmPRNum{216436} & 44 & 44 & 44 & 100.0 & 100.0 & 100.0 & 1.0 & 1.4 & 25.7 & 52 & 44 & 84.6 & 100.0 & 91.7 & 21.1 & 229 & 24.5 & 410 & 44.0 \\
    \llvmPRNum{216470} & 333 & 333 & 333 & 100.0 & 100.0 & 100.0 & 2.8 & 3.1 & 26.3 & 3,817 & 333 & 8.7 & 100.0 & 16.0 & 43.9 & 1,254 & 133.3 & 1,362 & 144.3 \\
    \midrule
    Avg$_{\text{micro}}$ & \multirow{2}{*}{3,861} & \multirow{2}{*}{3,861} & \multirow{2}{*}{3,861} & 100.0 & 100.0 & 100.0 & \multirow{2}{*}{1.3} & \multirow{2}{*}{1.6} & \multirow{2}{*}{26.2} & \multirow{2}{*}{89,984} & \multirow{2}{*}{3,861} & 4.3 & 100.0 & 8.2 & \multirow{2}{*}{32.5} & \multirow{2}{*}{306} & \multirow{2}{*}{42.3} & \multirow{2}{*}{390} & \multirow{2}{*}{54.7} \\
    Avg$_{\text{macro}}$ & & & & 100.0 & 100.0 & 100.0 & & & & & & 49.7 & 100.0 & 55.4 & & & & & \\
    \bottomrule
  \end{tabular}
  }
\end{table}

\subsection{RQ2: Accuracy}
\label{subsec:rq2}

We score the matches of \sys{} and \daggrep{} against the \gt{} on the two datasets.

\myparagraph{\corpusName{}}
As \cref{tbl:accuracy} shows,
for all \NumExactPRsWord{} rewrites, the patterns match
where the rewrite applies.
For \NumZeroGTPRsWord{} of these rewrites,
no function in \corpusName{} triggers the rewrite,
and \sys{} correctly reports no match.
For one of these \NumZeroGTPRsWord{} rewrites, \llvmPR{209322},
\daggrep{} cannot express the \code{knownNegative} constraint
and reports \DaggrepZeroGTP{} false positives.

\myparagraph{Test Suites}
The test suites consist of small, hand-written IR functions,
which exercise forms that compiled real-world code rarely produces.
On them, the patterns for the \NumPRsWord{} rewrites reach \TestPrecision{} precision
and \TestRecall{} recall:
no false positive and \TestFN{} false negative
among \TestG{} ground-truth functions.
The false negative is a vector multiplication by the non-splat constant
\code{<i32 3, i32 5>} in a test of \llvmPR{214376}.
The rewrite uses \cpp{m_CheckedInt},
which checks a custom C++ lambda predicate on every vector element.
\sys{} has no counterpart for it yet,
so the pattern states the predicate as a constraint on a constant,
which binds only to scalars and splat vectors.

\myparagraph{Comparison with \daggrep{}}
For each rewrite, we use the closest \daggrep{} pattern from \cref{subsec:rq1},
which drops the conditions that \daggrep{} cannot express.
% These patterns over-approximate.
On \corpusName{}, \daggrep{} reaches \DaggrepRecall{} recall
but only \DaggrepPrecision{} precision,
with \DaggrepFP{} false positives against \MainFP{} for \sys{}.
On the test suites, its precision is \DaggrepTestPrecision{},
against \TestPrecision{} for \sys{}.

\finding{\sys{} patterns match exactly where all \NumExactPRsWord{} rewrites apply.
\Daggrep{} cannot state most preconditions, so it reports many false positives.}

\subsection{RQ3: Efficiency of the Index}
\label{subsec:rq3}

We measure how many functions the index prunes, how long queries take with and without it, and what each part of the index contributes.

\myparagraph{Index Cost}
Building the index over the whole corpus takes \IndexBuildSec{} seconds
(\IndexBuildNoVFSec{} without value-tracking \facts{}),
so it can simply be rebuilt when the corpus changes.

\myparagraph{Pruning}
The index leaves few candidates.
Per pattern, a median of \MedFuncSurvive{} of \NumFunctions{} functions
(\MedFuncSurvivePct{}) remain after pruning,
in a median of \MedModSurvive{} modules (\MedModSurvivePct{}).
Even for the least selective pattern, \llvmPR{204927},
only \MaxFuncSurviveAbs{} functions (\MaxFuncSurvive{}) remain.
% and for the most selective.
% the \code{xor} case of \llvmPR{205383},
% only \MinFuncSurviveAbs{}.
% Since modules are loaded whole,
% the bytes of IR read are a better measure of cost:
% a query reads a median of \MedBytesSurvivePct{} of the corpus,
% and at most \MaxBytesSurvivePct{}.

\myparagraph{Query Time}
\Cref{tbl:accuracy} lists the time of each rewrite.
With the index,
every query we ran finished in \IndexedSecRangeLo{}--\IndexedSecRangeHi{}
seconds on \NumThreads{} threads, with a median of \IndexedSecMedian{} seconds.
Without it, a query scans all \NumFunctions{} functions
and takes a median of \FullScanSecMedian{} seconds,
for a median per-rewrite speedup of \MedIndexSpeedup{}
(\MinIndexSpeedup{} to \MaxIndexSpeedup{}).
In our runs, both returned identical matches for every rewrite, confirming \cref{cor:exact}.
\Daggrep{} scans every module on every query.
On the same \NumThreads{} threads,
its median query takes \DaggrepSecMedian{} seconds
(\DaggrepSecLo{}--\DaggrepSecHi{}).
The web application runs on fewer cores and fetches the IR of each candidate
over the network, so it depends on the index even more:
without the index, a single query may take over an hour, so we did not measure full scans there.

\myparagraph{Ablation}
We ran an ablation of abstract constraint evaluation on all patterns (column $T_{\text{NC}}$ of \cref{tbl:accuracy}).
Without evaluating constraints in the abstract domain over the \features{},
the median pattern keeps \MedNoConstrFuncs{} functions instead of
\MedFuncSurvive{} (\MedNoConstrFactor{}),
and \NumNoConstrWorse{} of the \NumPatterns{} patterns keep more functions.
On the command line, checking these extra candidates across \NumThreads{} local cores
raises the median time from \IndexedSecMedian{} to \MedSecNoConstr{} seconds.
In the web application (column Web of \cref{tbl:accuracy}),
extra candidates must be fetched over the network:
the median time grows from \WebFullSecMedian{} to \WebNoConstrSecMedian{} seconds,
and several rewrites download hundreds of megabytes more.
The gain is largest when a pattern's structure is unselective and its selectivity lies in its constraints,
as in \llvmPR{209322} (\code{udiv i?bits ?x, ?y:knownNegative}),
which structurally matches any \code{udiv} instruction and relies entirely on
\code{:knownNegative} to prune candidates:
without constraint evaluation, candidates jump from \WebFullCandsMax{} to \WebNoConstrCandsMax{},
downloaded data from \WebFullMBMax{} to \WebNoConstrMBMax{}~MB,
and query time from \WebFullSecMax{} to \WebNoConstrSecMax{} seconds (\WebNoConstrFactorMax{}).

\finding{The index leaves a median of \MedFuncSurvivePct{} of the functions
without losing a match,
so a query over \NumFunctionsShort{} functions on \NumThreads{} threads takes a median of
\IndexedSecMedian{} seconds, \MedIndexSpeedup{} faster than a full scan,
against \DaggrepSecMedian{} seconds for \daggrep{}.
The index matters more in the web application, where it avoids downloading the IR of most functions.}

% \subsection{Threats to Validity}
% \label{subsec:threats}

% Our folds are recent and may not represent older, larger folds;
% their variety (\cref{tbl:accuracy}) makes us expect similar results.
% The patterns were written by the authors,
% who know the language well;
% a user study is future work.
% The \gt{} comes from instrumenting each fold by hand;
% a mistake there would show up as a disagreement,
% and we traced every disagreement to its cause.

\section{Discussion}
\label{sec:discussion}

We discuss what \sys{} matches and how its design extends.

% %
% \begin{wrapfigure}{r}{0.4\linewidth}
% \vspace{-1em}
% \begin{lstlisting}[style=llvmstyle,basicstyle=\ttfamily\scriptsize]
% %t = trunc nsw i64 %iv to i32
% %p = phi i32 [ %a, %entry ],
%              [ %t, %loop ]
% %e = sext i32 %p to i64
% \end{lstlisting}
% \vspace{-0.8em}
% \caption{A loop in which \llvmPR{204089} applies.}
% \Description{A short LLVM IR loop in which a sext is applied to a phi whose incoming value is a trunc nsw.}
% \label{fig:pr204089}
% \end{wrapfigure}%
% %

\myparagraph{Interaction Between Rewrites}
\sys{} finds where a rewrite applies to a static IR snapshot,
whereas rewrites in a live compiler interact dynamically.
Within a single \instcombine{} pass,
earlier rewrites may transform matched instructions
(preempting a target rewrite)
or construct the exact IR
shape it requires (enabling the target rewrite).
The full \code{O2} pipeline compounds these interactions
by invoking \instcombine{} repeatedly as other passes transform the IR.
% The extreme case is \llvmPR{204089},
% which simplifies $\mathit{zext}(\mathit{trunc}\ \code{nuw}\ x)$
% and $\mathit{sext}(\mathit{trunc}\ \code{nsw}\ x)$ to $x$.
% Alone, it applies nowhere in \corpusName{},
% yet in \code{-O2} it applies in \PipeGPrOutlier{} functions,
% because other rewrites create its shape, mostly through \code{phi} and \code{select}.
% In the loop of \cref{fig:pr204089}, no \code{sext} of a \code{trunc} occurs.
% However, a \code{sext} of a \code{phi} equals a \code{phi} of \code{sext}s,
% \(
%   \code{sext}\,(\code{phi}\ A, B) = \code{phi}\ (\code{sext}\ A), (\code{sext}\ B).
% \)
% When \instcombine{} applies this rewrite to \code{\%e},
% it creates \code{sext i32 \%t to i64},
% which the rewrite of \llvmPR{204089} then simplifies to \code{\%iv}.
Nevertheless, static queries remain predictive:
measured against where the rewrites apply in \code{O2},
our patterns achieve \PipeMacroPrecision{} precision and \PipeMacroRecall{} recall
(macro-averaged across all \NumPRsWord{} rewrites).
Accounting for interactions within \instcombine{} itself is an interesting direction for future work:
one could migrate \instcombine{}'s rewrites into \sys{} to match patterns against the intermediate forms a function can take.
Representing the function as an e-graph~\cite{willsey_egg_2021}
could support this: the index could track the \features{} of
all equivalent forms within each e-class,
allowing our abstract interpretation to apply.

% \myparagraph{Cyclic Patterns}
% Patterns are acyclic, while IR can be cyclic through \code{phi} nodes.
% A loop-carried value is still expressible:
% bind the incoming value to a \metavar{} and require it to equal the \code{phi}
% with a constraint, as in \code{;; ?incoming == ?phi}.

\myparagraph{Generality}
The design is not specific to \llvm{}.
Matching patterns with constraints and indexing functions by their \features{} apply to other IRs.
In particular, the abstract interpretation can be reused,
since it needs only \sites{}, \facts{}, and \abstractFaces{} for the operators.

\section{Related Work}
\label{sec:related}

\myparagraph{Languages for Peephole Optimizations}
\alive{}~\cite{lopes_provably_2015} introduced a \dsl{} for \llvm{} peephole
optimizations written in \llvmir{} syntax with preconditions and verified with an SMT solver;
\aliveTwo{}~\cite{lopes_alive2_2021} validates pairs of concrete IR functions.
The \sys{} \dsl{} adopts the idea of IR-like patterns,
but asks where an optimization applies rather than whether a rewrite is sound.
Whereas \alive{} fixes instruction flags and comparison predicates
(and limits preconditions to a fixed set of analyses such as \code{isPowerOf2} and \code{MaskedValueIsZero}),
\sys{} lets patterns bind flags, predicates, and types as \metavars{} and refer to them in constraints.
GCC's \texttt{match.pd}~\cite{noauthor_match_nodate,noauthor_gccgccmatchpd_nodate},
MLIR's PDLL~\cite{noauthor_pdll_nodate} and pattern generation
language~\cite{attrot_pattern_2026},
and Cranelift's ISLE~\cite{cranelift_isle} serve similar matching and rewriting roles,
but target in-compiler rewriting rather than providing an indexed search engine.

\myparagraph{Pattern Matching and Indexing}
Tree pattern matchers for instruction selection can be generated by tools such as
iburg~\cite{fraser_engineering_1992}.
Theorem provers retrieve the instances of a term from large term sets with
discrimination trees and path indexing~\cite{mccune_experiments_1992}
and compile subsumption checks into code trees~\cite{voronkov_anatomy_1995};
databases filter graphs by indexed paths or frequent substructures before subgraph
matching~\cite{shasha_algorithmics_2002,yan_graph_2004};
code search tools index the syntactic features of a corpus to retrieve similar code~\cite{luan_aroma_2019},
match wildcard patterns over syntax trees~\cite{matute_syntactic_2024},
or run declarative queries over a database extracted from the code~\cite{github_codeql};
and Bonsai~\cite{root_bonsai_2026} prunes subtrees of a data structure
with conditions that interval analysis derives from the query and per-node metadata.
Code search relies on syntactic structure, and Bonsai covers numeric and geometric predicates.
\sys{} instead handles the constraints of compiler peephole optimizations
by abstract interpretation~\cite{cousot_abstract_1977} over recorded \facts{},
to prune functions while provably never discarding a match.

\section{Conclusion}
\label{sec:conclusion}

We presented \sys{}, a system for matching \llvm{} peephole
optimizations that combines a pattern \dsl{} with an indexed search engine.
Its single constraint evaluator runs both on concrete IR and, via sound
abstract interpretation, on function \features{},
enabling queries over \NumFunctionsShort{} functions in seconds.
\sys{} enables developers to interactively assess and refine candidate rewrites
at corpus scale before committing to a C++ implementation.

\section*{AI Use}
\label{sec:ai-use}

We used Claude Code as a coding assistant, mainly for code refactoring,
utility scripts, and enriching tool features.
The authors designed the core language, the matching algorithm, and the abstract
interpretation, and reviewed every AI-assisted change.
No AI tool is part of the matching pipeline.

\section*{Data-Availability Statement}
\label{sec:data}

We open-source \sys{} together with our evaluation results at~\cite{metaphir_artifact}.
The web application and its published index are also hosted online.

\bibliographystyle{ACM-Reference-Format}
\bibliography{references}

\appendix
\section{Proofs}
\label{app:proofs}

\subsection{Proof of \Cref{thm:lowering} (Lowering Preserves Matches)}
\label{app:proof-lowering}

\begin{restate}{thm:lowering}{Lowering preserves matches}
For every well-formed surface pattern $\pat$ and function $F$,
\[
  \Match(\pat, F) \;=\;
  \{\, \sigma|_{\MVars(\pat)} \mid \varnt \in \Lower(\pat),\;
  \sigma \in \Match(\varnt, F) \,\}.
\]
\end{restate}

The proof proceeds by induction over lowering steps.
We first define the measure that each step decreases,
then show that a single step preserves matches,
and finally combine the steps.

\myparagraph{Termination}
A \composite{} body can invoke only \composites{} defined before it,
so definitions are not recursive.
Let the \emph{depth} of a \composite{} be one more than the largest depth of
the \composites{} its branches invoke, and zero if they invoke none.
Lowering removes one abstraction at a time.
We write $\pat \Rightarrow \{\pat_1, \ldots, \pat_k\}$
when removing one abstraction from $\pat$ yields the patterns
$\pat_1, \ldots, \pat_k$,
such as one per branch of an inlined \composite{}.
Lowering $\pat$ then lowers each $\pat_i$ in turn,
so $\Lower(\pat) = \bigcup_i \Lower(\pat_i)$.
Each step either removes a \metaflag{}, which adds no invocation,
or replaces an invocation of depth $d$ by bodies whose invocations have depth
below $d$.
So each step decreases the pair of the multiset of invocation depths
and the number of \metaflags{},
ordered lexicographically with the multiset order first,
and lowering terminates.

\myparagraph{Preservation}
We show that
$\Match(\pat, F) = \bigcup_i \Match(\pat_i, F)|_{\MVars(\pat)}$
for every step, case by case.
\begin{itemize}[leftmargin=*,nosep]
  \item \emph{\Metaflags{}.}
  Lowering replaces a \metaflag{} by its body as a constraint at the flagged
  instruction,
  which is the \metaflag{}'s meaning (\cref{subsec:lowering}),
  and changes no node, so $\Match(\pat, F) = \Match(\pat_1, F)$.
  \item \emph{\Composite{} invocations.}
  Let $\pat$ contain $\lv = m_{\flagspec}\,(\vec a)$,
  and let $\pat_i$ inline the $i$-th branch $\beta_i$ of $m$ that is
  compatible with $\flagspec$:
  its parameters substituted by $\vec a$, its result node renamed to $\lv$,
  and its other \metavars{} renamed apart.
  By definition, the invocation matches iff some compatible branch matches,
  so the equation holds once three points are checked.
  \emph{Compatibility:}
  for a flag $\fl$ of $m$, a branch either passes $\fl$ on to its result
  node, where $\flagspec(\fl)$ then imposes exactly the invocation's condition,
  or fixes $\fl$ to a Boolean $b$.
  A fixed branch can match only if $\flagspec(\fl) = +$ and $b = \true$,
  or $\flagspec(\fl) = -$ and $b = \false$,
  or $\fl \notin \dom(\flagspec)$,
  or $\flagspec(\fl) = \mv$, in which case lowering adds the constraint
  $\mv \mathbin{==} b$, which holds exactly when $\sigma(\mv) = b$;
  so the dropped branches have no match,
  and the kept ones bind $\mv$ as the invocation's meaning does.
  \emph{Hygiene:}
  the renamed \metavars{} are disjoint from those of $\pat$ and of other
  invocations,
  so a match of $\pat_i$ may bind them freely,
  and restricting to $\MVars(\pat)$ removes them,
  as the invocation's meaning restricts to the \metavars{} of $\pat$.
  \emph{Or-self:}
  an invocation with \code{:o(N)} marks each inlined root with $\mko{o_N}$,
  so by \textsc{M-Self} the root also matches what $o_N$ matches,
  which is the meaning of \code{:o(N)} on an invocation.
  \item \emph{The \code{:c} mark on \composites{}.}
  Lowering yields $\pat_1$ and $\pat_2$ with the invocation's operands in
  either order,
  and a \code{:c} invocation matches iff it matches in one of the two orders.
\end{itemize}
Well-formedness keeps each step well-sorted,
since the arguments $\vec a$ have the sorts that its form requires.

\myparagraph{Induction}
By induction on this measure.
A pattern without \composite{} invocations,
and without \metaflags{} other than the marks $\mkc$ and $\mko{\cdot}$,
is already a \variant{},
so $\Lower(\pat) = \{\pat\}$ and the theorem holds by definition.
Otherwise, take a step $\pat \Rightarrow \{\pat_1, \ldots, \pat_k\}$,
and assume the theorem for each $\pat_i$.
Then
\begin{align*}
  \Match(\pat, F)
  &= \textstyle\bigcup_i \Match(\pat_i, F)|_{\MVars(\pat)}
  && \text{(one step)} \\
  &= \textstyle\bigcup_i \{\, \sigma|_{\MVars(\pat_i)}|_{\MVars(\pat)} \mid
    \varnt \in \Lower(\pat_i),\; \sigma \in \Match(\varnt, F) \,\}
  && \text{(hypothesis)} \\
  &= \{\, \sigma|_{\MVars(\pat)} \mid \varnt \in \Lower(\pat),\;
    \sigma \in \Match(\varnt, F) \,\}.
\end{align*}
The last step uses $\MVars(\pat) \subseteq \MVars(\pat_i)$,
since a step only adds renamed \metavars{} and constraints,
so the two restrictions amount to one,
and $\Lower(\pat) = \bigcup_i \Lower(\pat_i)$.
\qed

\subsection{Proofs of \Cref{lem:requirements,lem:typings}}
\label{app:proof-lemmas}

\begin{restate}{lem:requirements}{Requirements are necessary}
If $\Match(\varnt, F) \neq \emptyset$, then $\Req(\varnt) \subseteq \feat$.
\end{restate}

\begin{restate}{lem:typings}{Typings are sound}
If $\sigma \in \Match(\varnt, F)$,
then some $\theta \in \JoinSig(\varnt)$ satisfies
$\theta \subseteq \sigma$.
\end{restate}

These proofs and the proof of \cref{lem:eval-sound} use one fact.
Since a certain node $I$ has no $\mko{\cdot}$ mark,
$\pmatch{\sigma}{I}{\sigma(\lv_I)}$ is derived by \textsc{M-Inst}
or \textsc{M-Comm},
so the instruction $\sigma(\lv_I)$ that $\sigma$ maps $I$ to,
the \emph{image} of $I$, has the opcode of $I$.

\emph{\Cref{lem:requirements}.}
Each requirement is witnessed by $\sigma$:
a certain node's image has its written flags and lacks its negated ones;
a certain node with a certain operand node gives an edge between their images;
a \metavar{} used by two certain nodes is a shared operand of their images.

\emph{\Cref{lem:typings}.}
For each certain node,
the signature of its image is among the signatures of $F$
and is consistent with $\sigma$'s bindings of type-level \metavars{}.
The function $\JoinSig$ includes every consistent combination,
so it includes $\sigma$ restricted to the type-level \metavars{} of the certain
nodes, $\theta_\sigma \subseteq \sigma$.
If the set exceeds the cap,
the empty typing satisfies $\emptyset \subseteq \sigma$ trivially.
\qed

\subsection{Proof of \Cref{lem:eval-sound} (Abstract Evaluation Is Sound)}
\label{app:proof-soundness}

\begin{restate}{lem:eval-sound}{Abstract evaluation is sound}
Let $\sigma \in \Struct(\varnt, F)$ and $\theta \subseteq \sigma$.
For every constraint $e @ \lv_c$ of $\varnt$
and every subexpression $e'$ of $e$,
$\sem{e'}_\sigma \in \gam(\asem{e'}_\theta)$.
\end{restate}

Fix $\sigma \in \Struct(\varnt, F)$ and $\theta \subseteq \sigma$,
and recall that $\err \in \gam(\hat a)$ for every $\hat a$.
We prove $\sem{e}_\sigma \in \gam(\asem{e}_\theta)$ by induction on $e$.

\emph{Literals.}
We have $\sem{k}_\sigma = k \in \gam(\conc{k})$.

\emph{Variables.}
If $\mv \in \dom(\theta)$,
then $\sem{\mv}_\sigma = \sigma(\mv) = \theta(\mv) \in
\gam(\conc{\theta(\mv)})$.
Otherwise,
if $\SitesOf(\mv) = S \neq \emptyset$,
take any $s \in S$.
By the definition of $\SitesOf$,
$\mv$ is operand $j$ of a certain node $I$ with opcode $\op$.
By \cref{app:proof-lemmas},
the image of $I$ is an instruction with opcode $\op$.
Without the $\mkc$ mark, $s = (\op, \{j\})$,
and $\sigma(\mv)$ is operand $j$ of the image.
With the $\mkc$ mark, \textsc{M-Comm} may swap the operands,
so $\sigma(\mv)$ is operand $0$ or $1$ of the image,
and $s = (\op, \{0, 1\})$ by the definition of $\SitesOf$
(\cref{subsec:evaluator}).
A local name $\lv$ and its capture may also have the result \site{}
$s = (\op, \{\res\})$ of their own certain node $I$;
there, $\sigma(\lv)$ is the image of $I$ itself,
an instruction with opcode $\op$.
In every case, $\sigma(\mv) \in \ValsOf(s)$.
Since this holds for every $s \in S$,
$\sigma(\mv) \in \gam(\sitesv{S})$.
Otherwise the result is $\unk$.

\emph{Operators.}
Let $\hat a_i = \asem{e_i}_\theta$ and $w_i = \sem{e_i}_\sigma \in \gam(\hat a_i)$ by
induction.
If some $w_i = \err$,
then $\sem{f(\vec e)}_\sigma = \err$,
which lies in every concretization.
Otherwise:
(i) if every $\hat a_i = \conc{c_i}$, then $w_i = c_i$,
and $\sem{f(\vec e)}_\sigma = f(\vec c) = \asem{f(\vec e)}_\theta$,
or both are $\err$;
(ii) if $\hat{f}(\vec{\hat a}) = \conc{\false}$,
the soundness of \abstractFaces{} gives $f_{F,u}(\vec w) \in \{\false,
\err\} \subseteq \gam(\conc{\false})$;
(iii) otherwise the result is $\unk$.

\emph{Conjunction.}
Let $\hat a_1 = \asem{e_1}_\theta$ and $r_1 = \sem{e_1}_\sigma \in \gam(\hat a_1)$.
\begin{itemize}[leftmargin=*,nosep]
  \item $\hat a_1 \in \{\conc{\false}, \err\}$:
  then $r_1 \in \{\false,
  \err\}$,
  so $\sem{e_1 \mathbin{\&\&} e_2}_\sigma \in \{\false,
  \err\} \subseteq \gam(\hat a_1)$.
  \item $\hat a_1 = \conc{\true}$:
  then $r_1 \in \{\true,
  \err\}$,
  so the result is $\sem{e_2}_\sigma$ or $\err$,
  both in $\gam(\asem{e_2}_\theta)$ by induction.
  \item $\hat a_1 \notin \mathbb{B}$ and $\asem{e_2}_\theta \in \{\conc{\false}, \err\}$:
  by induction, $\sem{e_2}_\sigma \in \{\false, \err\}$.
  In concrete evaluation, $e_1 \mathbin{\&\&} e_2$ either short-circuits to
  $\false$ (when $r_1 = \false$), evaluates to $\sem{e_2}_\sigma \in \{\false,
  \err\}$ (when $r_1 = \true$), or fails with $\err$;
  in all cases, $\sem{e_1 \mathbin{\&\&} e_2}_\sigma \in \{\false, \err\} =
  \gam(\conc{\false})$.
  \item Otherwise, $\asem{e_1 \mathbin{\&\&} e_2}_\theta = \unk$, and $\gam(\unk) = \CVal_\err$.
\end{itemize}

\emph{Disjunction} is dual, with $\conc{\true}$ deciding.
When the left side is unknown and the right side is $\conc{\false}$ or $\err$,
the result is $\unk$,
which is trivially sound.

\emph{Negation.}
If $\asem{e}_\theta = \conc{b}$ for a Boolean $b$,
then $\sem{e}_\sigma \in \{b,
\err\}$,
so $\sem{{!}e}_\sigma \in \{\neg b, \err\} = \gam(\conc{\neg b})$;
if $\asem{e}_\theta = \err$,
the result is $\err$;
otherwise $\unk$.
\qed

\section{The Matching Algorithm}
\label{app:algorithm}

\Cref{alg:match} computes $\Match(\varnt, F)$.
It matches the roots of $\varnt$ one after another
(line~\ref{ln:roots}),
each against every instruction of $F$ (line~\ref{ln:every-inst}),
and descends from each node to its operands (line~\ref{ln:operands}).
Matching extends a partial \instantiation{}:
an \emph{extension} of $\sigma$ keeps every binding of $\sigma$
and binds more names.
\textsc{MatchNode} tries each rule of \cref{fig:match-rules} that applies
(line~\ref{ln:rules}),
so the search explores every choice the rules allow.
A node whose local name is an operand of two other nodes is reached twice.
The second time, \textsc{MatchNode} only checks that $\sigma(\lv)$
is the same value (line~\ref{ln:shared}),
so both uses refer to one instruction.
After the roots, $\Sigma$ is $\Struct(\varnt, F)$,
which \textsc{Match} then filters by \cref{def:match} (line~\ref{ln:filter}).

\begin{algorithm}[t]
\footnotesize
\caption{Matching a \Variant{} $\varnt$ against a Function $F$.}
\label{alg:match}
\SetKwFunction{MatchVariant}{Match}
\SetKwFunction{MatchNode}{MatchNode}
\SetKwFunction{MatchOperand}{MatchOperand}
\SetKwProg{Fn}{function}{}{}
\SetKwComment{tcp}{// }{}
\Fn{\MatchVariant{$\varnt$, $F$}}{
  $\Sigma \gets \{\emptyset\}$ \tcp*{partial \instantiations{}}
  \ForEach(\tcp*[f]{assume roots have no $\mko{\cdot}$ mark}){root $I$ of $\varnt$\nllabel{ln:roots}}{
    $\Sigma' \gets \emptyset$ \tcp*{extensions that also match $I$}
    \ForEach{$\sigma$ in $\Sigma$}{
      \lForEach{instruction $u$ of $F$}{
        $\Sigma' \gets \Sigma' \cup \MatchNode{$I$, $u$, $\sigma$}$\nllabel{ln:every-inst}
      }
    }
    $\Sigma \gets \Sigma'$ \tcp*{each $\sigma$ replaced by its extensions}
  }
  \Return{$\{\sigma \in \Sigma \mid \sem{e}_\sigma = \true
  \text{ for every constraint } e @ \lv \text{ of } \varnt\}$}\nllabel{ln:filter}
  \tcp*{\cref{def:match}}
}
\Fn(\tcp*[f]{every way that node $I_\lv$ matches $u$}){\MatchNode{$I_\lv$, $u$, $\sigma$}}{
  \lIf{$\sigma(\lv) = u$\nllabel{ln:shared}}{\Return{$\{\sigma\}$}}
  \lElseIf{$\sigma(\lv)$ is defined}{\Return{$\emptyset$}}
  $R \gets \emptyset$\;
  \ForEach{rule $\rho$ of \textsc{M-Inst}, \textsc{M-Comm}, \textsc{M-Self}
  that applies to $I_\lv$\nllabel{ln:rules}}{
    \If(\tcp*[f]{operands are matched below}){$u$ satisfies the premises of
    $\rho$ other than the operand matches}{
      $S \gets \{\sigma'\}$, where $\sigma'$ extends $\sigma[\lv \mapsto u]$
      with the bindings of these premises\;
      \ForEach{operand $o$ of $I_\lv$ and the value $u'$ that $\rho$ matches it
      against\nllabel{ln:operands}}{
        $S' \gets \emptyset$\;
        \lForEach{$\sigma''$ in $S$}{
          $S' \gets S' \cup \MatchOperand{$o$, $u'$, $\sigma''$}$
        }
        $S \gets S'$\;
      }
      $R \gets R \cup S$\;
    }
  }
  \Return{$R$}\;
}
\Fn{\MatchOperand{$o$, $u$, $\sigma$}}{
  \lIf{$o$ is a local name $\lv$}{\Return{\MatchNode{$I_\lv$, $u$, $\sigma$}}}
  \lElseIf{$o$ is a \metavar{} $\mv$ that $u$ fits by sort,
  and $\sigma(\mv)$ is undefined or $\sigma(\mv) \doteq u$}{\Return{$\{\sigma[\mv \mapsto u]\}$}}
  \lElseIf{$o$ is a constant and $o \doteq u$}{\Return{$\{\sigma\}$}}
  \Return{$\emptyset$}\;
}
\end{algorithm}

\section{Patterns of the Evaluation}
\label{app:expressiveness}

\Cref{tbl:expressiveness} compares the \sys{} pattern of each rewrite
with the closest \daggrep{} patterns (\cref{subsec:rq1}),
and \cref{fig:pr216470} shows the largest \sys{} pattern
next to the C++ code it describes.

\begin{table}[h]
  \centering
  \caption{Per-Rewrite Patterns in \sys{} and in \daggrep{}.
  Inst and Cons count the instruction and constraint lines of the \sys{} patterns
  (summing two patterns each for \llvmPR{205383}, \llvmPR{207462}, and \llvmPR{212132});
  Files counts the \daggrep{} patterns written for the rewrite, and
  a check mark means that \daggrep{} matches the same instruction structure as \sys{}.
  The last column lists the conditions that \daggrep{} cannot state and that we drop.}
  \label{tbl:expressiveness}
  \footnotesize
  \begin{tabular}{@{}rrrrrcp{0.5\linewidth}@{}}
    \toprule
    & \multicolumn{2}{c}{\textbf{\sys{}}} & \multicolumn{3}{c}{\textbf{\daggrep{}}} & \\
    \cmidrule(lr){2-3}\cmidrule(lr){4-6}
    \textbf{PR} & \textbf{Inst} & \textbf{Cons} & \textbf{Files} & \textbf{Inst} & \textbf{Exact} & \textbf{Dropped for \daggrep{}} \\
    \midrule
    \llvmPRNum{204927} & 3 & 0 & 3 & 9 &  & single use \\
    \llvmPRNum{204928} & 3 & 0 & 3 & 9 &  & single use \\
    \llvmPRNum{205383} & 10 & 3 & 4 & 24 &  & bit-width equations; predicate \\
    \llvmPRNum{207462} & 7 & 4 & 2 & 7 &  & single use; \hcode{?c0 != 0}; \hcode{notUndef} \\
    \llvmPRNum{207564} & 3 & 0 & 1 & 3 & \checkmark & -- \\
    \llvmPRNum{209322} & 1 & 0 & 1 & 1 &  & \hcode{knownNegative} \\
    \llvmPRNum{210499} & 2 & 1 & 4 & 8 &  & the \hcode{i1} result type \\
    \llvmPRNum{212132} & 12 & 0 & 2 & 12 &  & \hcode{knownPowerOfTwo}; single use; predicate \\
    \llvmPRNum{213313} & 5 & 0 & 2 & 6 & \checkmark & -- \\
    \llvmPRNum{213709} & 3 & 4 & 2 & 6 &  & \hcode{sgt(?bits, 1)}; predicate; single use \\
    \llvmPRNum{214376} & 2 & 0 & 1 & 2 &  & odd constant (any constant) \\
    \llvmPRNum{214381} & 3 & 2 & 1 & 3 &  & \hcode{?d != 0}; predicate and constant ranges \\
    \llvmPRNum{214517} & 2 & 0 & 1 & 2 &  & single use \\
    \llvmPRNum{214541} & 2 & 1 & 1 & 2 &  & \hcode{noOverflow}; single use \\
    \llvmPRNum{215122} & 3 & 0 & 1 & 3 &  & \hcode{lowBitMask} (any constant); single use \\
    \llvmPRNum{215130} & 2 & 1 & 1 & 2 &  & \hcode{?c + ?negc == 0}; single use \\
    \llvmPRNum{216046} & 1 & 1 & 1 & 1 &  & \hcode{inSignedRange} \\
    \llvmPRNum{216436} & 3 & 0 & 1 & 3 &  & \hcode{lowBitMask} (any constant); single use \\
    \llvmPRNum{216470} & 6 & 2 & 2 & 11 &  & \hcode{lowBitMask}, \hcode{popcount}, \hcode{noOverflow}; single use; predicate \\
    \llvmPRNum{204089} & 2 & 1 & 2 & 4 & \checkmark & -- \\
    \bottomrule
  \end{tabular}
\end{table}

\begin{figure}[h]
  \centering
  \begin{subfigure}[t]{0.43\textwidth}
\begin{lstlisting}[style=cppstyle,basicstyle=\scriptsize\ttfamily,xleftmargin=0pt]
match(&I, m_c_Add(
  m_ZExt(m_SpecificICmp(ICmpInst::ICMP_NE,
    m_And(m_Value(X), m_LowBitMask(Mask)),
    m_ZeroInt())),
  m_OneUse(m_ZExtOrSelf(m_OneUse(
    m_LShr(m_Deferred(X), m_APInt(ShAmt)))))))
&& Mask->popcount() == *ShAmt
&& (NeedZext ||
    willNotOverflowUnsignedAdd(X, MaskC, I))
\end{lstlisting}
    \caption{C++ Rewrite in \instcombine{}.}
  \end{subfigure}
  \hfill
  \begin{subfigure}[t]{0.55\textwidth}
\begin{lstlisting}[style=mpirstyle,basicstyle=\scriptsize\ttfamily,xleftmargin=0pt]
%and = and:c i?bits ?x, ?mask:lowBitMask
%cmp = icmp ne i?bits %and, 0
%ext = zext i1 %cmp to ?rty
%shr = lshr:s i?bits ?x, ?shift:apIntAllowPoison
%zext = zext:?zxt:s:o i?bits %shr to ?rty
%add = add:c ?rty %ext, %zext
;; popcount(?mask) == ?shift
;; opcode(?zxt) == "zext" || noOverflow("uadd", ?x, ?mask)
\end{lstlisting}
    \caption{\sys{} Pattern.}
  \end{subfigure}
  \Description{Two code listings side by side: the C++ PatternMatch of the rewrite and the equivalent MetaphIR pattern.}
  \caption{The Largest Pattern of the Evaluation, for \llvmPR{216470}:
  $(X \shrop C) + \mathit{zext}((X \mathbin{\&} M) \neq 0)$ with $M = 2^C - 1$,
  a Rounding-up Division by $2^C$.
  The operator $\shrop$ is a logical right shift.
  The capture \code{:?zxt} binds the matched \code{zext}, or the \code{lshr} when \code{:o} skips the \code{zext},
  so the last constraint applies the no-overflow check only when there is no \code{zext},
  like \cpp{NeedZext} in the C++ code.}
  \label{fig:pr216470}
\end{figure}

\end{document}